\documentclass{IEEEtran}
\usepackage{cite}
\usepackage{amsmath,amssymb,amsfonts}
\usepackage{algorithmic}
\usepackage{graphicx}
\usepackage{textcomp}
\def\BibTeX{{\rm B\kern-.05em{\sc i\kern-.025em b}\kern-.08em
    T\kern-.1667em\lower.7ex\hbox{E}\kern-.125emX}}
    
\usepackage{overpic}
\usepackage{mathtools}
\usepackage{tikz}
\usetikzlibrary{calc,positioning}
\usepackage{array}
\usepackage[caption=false,font=footnotesize]{subfig}
\usepackage{siunitx}
\usepackage[crop=pdfcrop]{pstool}
\usepackage{overpic}
\usepackage{cancel}
\usepackage{mwe} 
\usepackage{bm}
\usepackage{dblfloatfix} 
\usepackage{tabularx}
\usepackage[usestackEOL]{stackengine}
\usepackage{blindtext}
\usepackage{float}
\usepackage{flushend}

\makeatletter
\def\user@resume{resume}
\def\user@intermezzo{intermezzo}
\newcounter{previousequation}
\newcounter{lastsubequation}
\newcounter{savedparentequation}
\renewenvironment{subequations}[1][]{%
      \def\user@decides{#1}%
      \setcounter{previousequation}{\value{equation}}%
      \ifx\user@decides\user@resume 
           \setcounter{equation}{\value{savedparentequation}}%
      \else  
      \ifx\user@decides\user@intermezzo
           \refstepcounter{equation}%
      \else
           \setcounter{lastsubequation}{0}%
           \refstepcounter{equation}%
      \fi\fi
      \protected@edef\theHparentequation{%
          \@ifundefined {theHequation}\theequation \theHequation}%
      \protected@edef\theparentequation{\theequation}%
      \setcounter{parentequation}{\value{equation}}%
      \ifx\user@decides\user@resume 
           \setcounter{equation}{\value{lastsubequation}}%
         \else
           \setcounter{equation}{0}%
      \fi
      \def\theequation  {\theparentequation  \alph{equation}}%
      \def\theHequation {\theHparentequation \alph{equation}}%
      \ignorespaces
}{%
  \ifx\user@decides\user@resume
       \setcounter{lastsubequation}{\value{equation}}%
       \setcounter{equation}{\value{previousequation}}%
  \else
  \ifx\user@decides\user@intermezzo
       \setcounter{equation}{\value{parentequation}}%
  \else
       \setcounter{lastsubequation}{\value{equation}}%
       \setcounter{savedparentequation}{\value{parentequation}}%
       \setcounter{equation}{\value{parentequation}}%
  \fi\fi
  \ignorespacesafterend
}
\makeatother

\begin{document}

\newcommand{\tensor}[1]{\overline{\overline{#1}}}

\newcommand{\nullv}{\varnothing}

\newcommand{\xv}{\bm{\hat{x}}}
\newcommand{\yv}{\bm{\hat{y}}}
\newcommand{\zv}{\bm{\hat{z}}}

\newcommand{\Mv}{\bm{M}}
\newcommand{\Pv}{\bm{P}}
\newcommand{\Qv}{\bm{Q}}

\newcommand{\Hiv}{\bm{H}_i}
\newcommand{\Hrv}{\bm{H}_r}
\newcommand{\Htv}{\bm{H}_t}

\newcommand{\Eiv}{\bm{E}_i}
\newcommand{\Erv}{\bm{E}_r}
\newcommand{\Etv}{\bm{E}_t}

\newcommand{\Ev}{\bm{E}}
\newcommand{\Hv}{\bm{H}}

\newcommand{\DEv}{\Delta\bm{E}}
\newcommand{\DHv}{\Delta\bm{H}}

\newcommand{\Eavv}{\bm{E}_\text{av}}
\newcommand{\Havv}{\bm{H}_\text{av}}

\newcommand{\Eav}{E_\text{av}}
\newcommand{\Hav}{H_\text{av}}

\newcommand{\htext}[1]{%
	\makebox[0pt]{\Centerstack{#1}}
}

\newcommand{\vtext}[1]{%
	\makebox[0pt]{\rotatebox[origin=c]{90}{\Centerstack{#1}}}
}

\makeatletter
\newcommand{\subalign}[1]{%
  \vcenter{%
    \Let@ \restore@math@cr \default@tag
    \baselineskip\fontdimen10 \scriptfont\tw@
    \advance\baselineskip\fontdimen12 \scriptfont\tw@
    \lineskip\thr@@\fontdimen8 \scriptfont\thr@@
    \lineskiplimit\lineskip
    \ialign{\hfil$\m@th\scriptstyle##$&$\m@th\scriptstyle{}##$\hfil\crcr
      #1\crcr
    }%
  }%
}
\makeatother

\title{Floquet Analysis of Space-Time Modulated Huygens' Metasurfaces with Lorentz Dispersion}

\author{Ville Tiukuvaara, \IEEEmembership{Student Member, IEEE}, Tom J. Smy, and Shulabh Gupta, \IEEEmembership{Senior Member, IEEE}

\thanks{Ville Tiukuvaara, Tom J. Smy, and Shulabh Gupta are with Carleton University, Ottawa, Canada (e-mail: villetiukuvaara@cmail.carleton.ca). }
}

\maketitle

\begin{abstract}
A rigorous semi-analytical Floquet analysis is proposed for a zero-thickness space-time modulated Huygens' metasurface to model and determine the strengths of the new harmonic components of the scattered fields. The proposed method is based on Generalized Sheet Transition Conditions (GSTCs) treating a metasurface as a spatial discontinuity. The metasurface is described in terms of Lorentzian electric and magnetic surface susceptibilities, $\chi_\text{e}$ and $\chi_\text{m}$, respectively, with parameters (e.g. resonant frequency) that are periodically modulated in both space and time. The unknown scattered fields are expressed in terms of Floquet harmonics, for which the amplitudes can be found by numerically solving a set of linear equations, leading to the total scattered fields. Using existing computational techniques, the method is validated using several examples of pure-space and pure-time modulation with different modulation strengths and pumping frequencies. Finally, two cases of space-time modulation (standing wave perturbation and a travelling wave perturbation) are presented to demonstrate the breaking of Lorentz reciprocity. The proposed method is simple and versatile and able to determine the steady-state response of a space-time modulated Huygen's metasurface that is excited with an oblique plane wave, or a general incident field such as a Gaussian beam.
\end{abstract}


\section{Introduction}

Space-time modulated materials were studied in the context of parametric amplification \cite{Cullen:1958aa}\cite{Cassedy:1963aa} in the 1950s, and have received renewed interest recently in the context of metamaterials. While static metamaterials have provided a plethora of wave manipulation devices \cite{Chen:2016aa}\cite{Genevet:2015aa}, they are typically limited by Lorentz reciprocity. While this can be overcome through the use of magnetic-optic materials \cite{Adam:2002aa} or nonlinear materials \cite{Shi:2015aa}, these methods require bulky implementations or provide weak non-reciprocity, respectively. For this reason, \textit{space-time modulated} metamaterials have emerged as an appealing alternative: by modulating the constitutive parameters of a linear medium in space and time, it is possible to achieve strong reciprocity. This has been explored in bulk metamaterials \cite{Caloz:2020aa}\cite{Caloz:2020ab} and metasurfaces \cite{Wang:2020aa}\cite{Hadad:2015aa} with applications including isolators, circulators, and frequency mixers \cite{Taravati:2017aa,Taravati:2020aa,Ramaccia:2020aa}, and the possibility of using space-time diffraction patterns as channels for wireless communications \cite{Taravati:2019aa}.

At the same time, there is a strong interest in Huygen's metasurfaces due to their impedance matching capabilities with free-space and their versatile applications in wavefront shaping \cite{Elliptical_DMS,GeneralizedRefraction,meta3}. They are constructed using a 2-D array of electrically small Huygen's sources, exhibiting perfect cancellation of backscattered fields, due to optimal interactions of their electric and magnetic dipolar moments \cite{Kerker_Scattering}. Some efficient implementations of Huygens' metasurfaces are based on all-dielectric resonators \cite{Kivshar_Alldielectric}\cite{Elliptical_DMS}\cite{AllDieelctricMTMS} and orthogonally collocated small electric and magnetic dipoles \cite{Grbic_Metasurfaces}\cite{HuygenBook_Eleftheriades}.

Consequently, a combination of the wave-shaping capabilities of Huygens' metasurfaces with space-time modulation principles, is an interesting avenue to explore for advanced electromagnetic wave control, in both space and time. To investigate into the properties of space-time modulated Huygens' metasurfaces, finite-difference time-domain (FDTD) techniques have recently been proposed to analyze a zero thickness model of Huygens' metasurfaces \cite{Smy:2017aa,Stewart:2018aa,Smy:2020aa}, based on the the generalized sheet transition conditions (GSTCs) \cite{Kuester:2003aa}. Unlike frequency domain techniques that are typically used for static metasurfaces like the finite-difference frequency-domain (FDFD) method \cite{Vahabzadeh:2018ab} and the boundary element method (BEM) \cite{Stewart:2019aa}, FDTD lends itself naturally when the surface is time-varying. However, if the modulation and incident field are periodic, then a steady-state will be achieved that is inefficient to compute with FDTD. For such \textit{space-time periodic} metasurfaces, it is desirable to have an efficient method for computing the steady-state scattered fields, which are expressed in the form of space-time Floquet harmonics.

An important aspect to consider in the analysis of space-time modulated metasurfaces is that of \emph{temporal dispersion}. Given metasurfaces are constructed using sub-wavelength resonators which are inherently dispersive, their corresponding surface susceptibilities are naturally frequency dependent. Moreover, quite often the surface is operated near resonance for maximal interaction of the waves with the surface. Since time modulation leads to generation of new temporal frequency components different from those of the excitation time-domain signal, incorporating metasurface dispersion in the space-time analysis is important for obtaining correct field solutions. Keeping causality of the surface in mind, frequency dependent surface susceptibility distribution may not be arbitrarily applied and must be chosen with care.

Several methods have recently been presented. It is possible to model a metasurface using surface impedances as in \cite{Wang:2020aa}\cite{Hadad:2015aa}, but these works concentrate on ``travelling wave'' space-time modulations exclusively. 
Several other methods have been shown that treat the metasurface as a finite-thickness medium \cite{Taravati:2019aa}\cite{Inampudi:2019aa}, but the modelling of a bulk medium adds unnecessary computational burden if the metasurface can be treated as a zero-thickness sheet. In this work, we treat the surface as such, using surface susceptibilities following a physically motivated Lorentzian profile to account for temporal dispersion, whose parameters (e.g. resonant frequency) are parametrized to emulate a space-time modulation of the metasurface. Combined with GSTCs, the Floquet harmonic amplitudes are computed by solving a set of linear equations. The proposed semi-analytical method thus efficiently computes the steady-state response of a zero-thickness space-time modulated Huygens' metasurface that is excited by a plane wave. We also show how the method can be extended to arbitrary excitations, such as Gaussian beams, by decomposing such fields into plane waves using Fourier decomposition.

The paper is structured as follows. Section~\ref{Sec:Overview} describes the problem statement of this work, and provides background on time-varying metasurfaces and how these can be modelled with Lorentzian susceptibilities. Section~\ref{Sec:Formulation} presents the proposed method based on an expansion using Floquet harmonics, forming a set of linear equations that can be solved for the harmonic amplitudes and used to construct the fields. Examples are provided in Section~\ref{Sec:Results} for cases of pure-space and pure-time modulation with comparison to FDFD and FDTD to validate the method, and followed by two types of space-time modulation to demonstrate violating Lorentz reciprocity. Finally, conclusions are provided in Section~\ref{Sec:Conclusion}.

\section{Space-Time Modulated Metasurfaces}\label{Sec:Overview}

\subsection{Problem Statement}

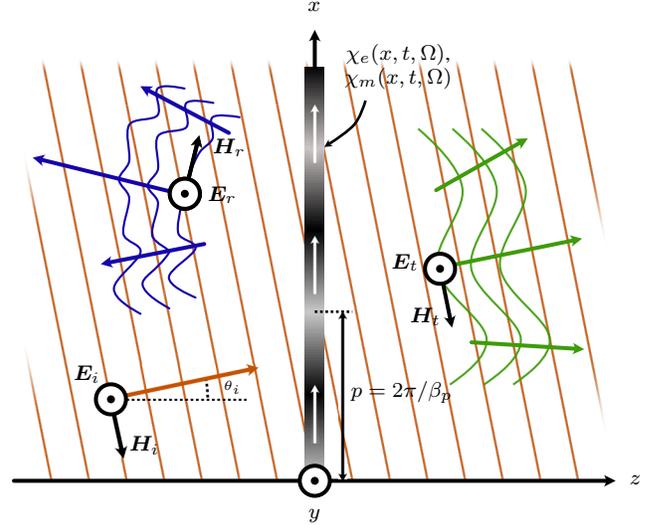
\begin{figure}[h]
	\begin{overpic}[grid=false,trim={0cm 0 0cm 0},clip]{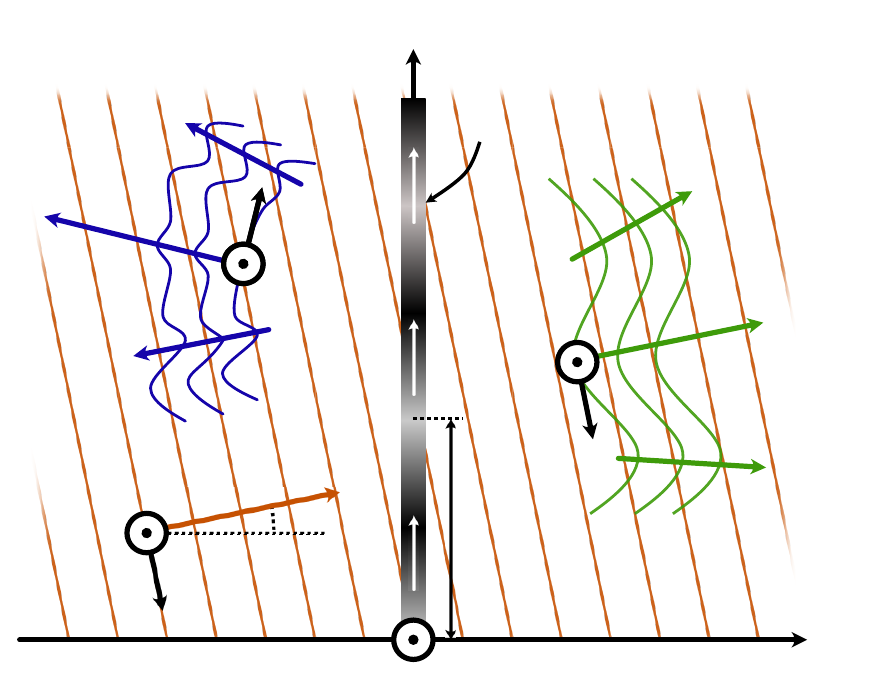}
		\put(94,5){\footnotesize \htext{$z$}}
		\put(46.6,0){\footnotesize \htext{$y$}}
		\put(46.6,75){\footnotesize \htext{$x$}}
		\put(59,66){\footnotesize \htext{$\chi_{e}(x,t,\Omega)$,\\$\chi_{m}(x,t,\Omega)$}}
		\put(52,18){\footnotesize $p=2\pi/\beta_p$}
		\put(13,20.5){\footnotesize \htext{$\Eiv$}}
		\put(21.5,10){\footnotesize \htext{$\Hiv$}}
		\put(34.5,19){\tiny \htext{$\theta_i$}}
		\put(60,37){\footnotesize \htext{$\Etv$}}
		\put(63,29){\footnotesize \htext{$\Htv$}}
		\put(33,47){\footnotesize \htext{$\Erv$}}
		\put(34,54){\footnotesize \htext{$\Hrv$}}
	\end{overpic}
	\caption{An illustration of scattering from from a space-time modulated metasurface with a spatial period $p$, where at TE plane wave incident at an angle $\theta_i$ produces reflected and transmitted fields. The metasurface has zero thickness and is infinite in size.}
	\label{Fig:Problem}
\end{figure}

Consider the problem in Fig.~\ref{Fig:Problem}, where a metasurface placed at $z=0$ acts as a scatterer, producing reflected and transmitted fields, $\Erv$ and $\Etv$, when a incident field $\Eiv$ is present\footnote{In the absence of a metasurface, the incident field defined to be present everywhere, so the total field with the metasurface is $\Ev=\Eiv+\begin{cases}\Etv,z>0\\\Erv,z<0\end{cases}.$}. The local electric field at the metasurface induces electric and magnetic polarizations, $\Pv$ and $\Mv$. If there is no time modulation, the constitutive relations can be written in the frequency domain as \cite{Lindell:1994aa}\cite{Achouri:2015aa}
\begin{subequations}\label{Eq:PolarizationChiQM}
	\begin{equation}\label{Eq:PolarizationChiQ}
		\mathbf{P}(x,\omega) = \epsilon_0\tensor{\chi}_{ee}\mathbf{E}_{\text{av}}(x,\omega) + \tensor{\chi}_{em}\sqrt{\mu_0\epsilon_0}\mathbf{H}_{\text{av}}(x,\omega),%
	\end{equation}
	\begin{equation}\label{Eq:PolarizationChiM}
		\mathbf{M}(x,\omega) = \tensor{\chi}_{mm}\mathbf{H}_{\text{av}}(x,\omega)+ \tensor{\chi}_{me}\sqrt{\epsilon_0\mu_0}\mathbf{E}_{\text{av}}(x,\omega),%
	\end{equation}
\end{subequations}
where the average indicates the average of the total fields at $z=0$ in terms of fields at $z=0^-$ and $z=0^+$. %
%

The normal component of the susceptibility tensors $\tensor{\chi}$ are often zero or negligible \cite{Albooyeh:2017aa}\cite{Selvanayagam:2013aa}, and are discarded in this work, for simplicity. Furthermore, we will simplify to a uniaxial surface with no cross-coupling, so the equations simplify and can be written in the time domain as \cite{Lathi:2018aa}
\begin{subequations}
	\begin{align}
	Q_y(x,t) = \int_{-\infty}^\infty \chi_e(x,\tau)E_{\text{av},y}(x,t-\tau)\,d\tau\\
	M_x(x,t) = \int_{-\infty}^\infty \chi_m(x,\tau)H_{\text{av},y}(x,t-\tau)\,d\tau
	\end{align}
	\label{Eq:QMNoMod}
\end{subequations}
\noindent where the electric polarization has been normalized so $Q_y=P_y/\epsilon_0$. The averages of the fields at $z=0$ is
\begin{align}
	E_{\text{av},y} &= \left[ E_i+\frac{E_r+E_t}{2} \right],
	& H_{\text{av},x} &= \left[ H_{i,x}+\frac{H_{r,x}+H_{t,x}}{2} \right]. \label{Eq:AverageFields}
\end{align}
The transverse fields are also governed by the GSTC equations \cite{Achouri:2015aa}\cite{Idemen:1987aa}\cite{Kuester:2003aa}
\begin{subequations}\label{Eq:GSTC}
	\begin{align}
		\DEv(x,t)\times \zv &= \mu_0\frac{d\Mv(x,t)}{dt}-\nabla \cancelto{0}{Q_z} \times\zv,\\
		\zv\times \DHv(x,t) &= \epsilon_0\frac{d\mathbf{Q}(x,t)}{dt}-\zv\times\nabla \cancelto{0}{M_z},
	\end{align}
\end{subequations}
which in the TE case simplify to
\begin{subequations}
	\begin{align}
		E_t-E_r &= \mu_0 \frac{dM_x(x,t)}{dt},\label{Eq:GSTCM}\\
		H_{t,x}-H_{r,x} &=  \epsilon_0\frac{dQ_y(x,t)}{dt}.\label{Eq:GSTCQ}
	\end{align}\label{Eq:GSTCScalar}
\end{subequations}

Together, \eqref{Eq:QMNoMod} and \eqref{Eq:GSTCScalar} are sufficient to uniquely solve for the fields and polarizations, given $\chi_\text{e}$, $\chi_\text{m}$, and an incident field. We are interested in solving this problem, in the case that $\chi_\text{e}$ and $\chi_\text{m}$ are modulated in time in addition to space. Specifically, we will consider the form of the fields when the spatial variation is periodic (with spatial angular frequency $\beta_p$ as in Fig.~\ref{Fig:Problem}) and the time variation is periodic (with ``pumping'' frequency $\omega_p$). The resulting fields will be described as a summation of infinite space-time harmonics assuming a temporally dispersive metasurface, which we wish to determine.


\subsection{Linear Time-Variant (LTV) Systems}

We now consider how \eqref{Eq:QMNoMod} generalizes for time-dependent susceptibilties. This convolution can be viewed as an input (field)-output (polarization) system, which is linear and time-invariant (LTI). A time-varying susceptibility on the other hand represents a linear time-variant (LTV) system, where the electric polarization density can be written as \cite{Claasen:1982aa}\cite{Franks:1969aa} 
\begin{align}
	Q_y(x,t) = \int_{-\infty}^\infty \chi_{e}(x,t,\tau)\Eav(x,t-\tau)\,d\tau .\label{Eq:ImpulseResponseConvTime}
\end{align} 
This is a generalized convolution, where the impulse response $\chi_\text{e}(t,\tau)$ gives the response that is probed at time $t$ due to an input that is applied $t$ time units earlier. It is worth emphasizing that although there is time variation, this should not be confused with a nonlinear system; $\chi_\text{e}$ does not depend on the magnitude of $E_\text{av}$. Using the Fourier transform of $\chi_e$ with respect to $\tau$, this can be written for an arbitrary signal $E_\text{av}$ in an equivalent relation \cite{Claasen:1982aa}\cite{Zadeh:1950aa}
%
	\begin{align}
		%
		%
		Q_y(x,t) &= \frac{1}{2\pi}\int_{-\infty}^\infty \chi_{e}(x,t,\Omega)\Eav(x,\Omega)e^{j\Omega t}\,d\Omega. \label{Eq:LTVFreq}
	\end{align}
We observe that if the surface is time-invariant (no depenency on $t$), then neglecting the temporal dispersion is reasonable if $\Eav(\Omega)$ is monochromatic. However, even with a monochromatic incident field $E_i$, any time-dependence of the susceptibility will result in $Q_y(x,t)$ that is not monochromatic, and subsequently fields that are not monochromatic via \eqref{Eq:GSTCScalar}. Thus, the frequency dispersion inherent to a static metasurface must in general be considered when time modulation is added.

\subsection{Periodic Lorentzian Susceptibilities}


To model the temporal dispersion inherent to a static metasurface, a Lorentzian distribution provides a physically-motivated response that can be used to model metasurfaces, such as Huygens' metasurfaces \cite{Smy:2017aa}\cite[p.~317-318]{Rothwell:2018aa}. In the time-domain, this is a damped oscillator model that can be expressed
\begin{subequations}
	\begin{align}
		\left[\omega_{e0}^2(t)+\frac{d^2}{dt^2} + \alpha_{e}(t) \frac{d}{dt}\right]Q_y(x,t)&= \omega_{ep}^2(t) E_{\text{av},y}(x,t), \label{Eq:LorentzianQ}
	\end{align}
	\begin{align}
		\left[\omega_{m0}^2(t)+\frac{d^2}{dt^2} + \alpha_{m}(t) \frac{d}{dt}\right]M_x(x,t)&= \omega_{mp}^2(t) H_{\text{av},x}(x,t), \label{Eq:LorentzianM}
	\end{align}\label{Eq:Lorentzian}
\end{subequations}%

\noindent where $\omega_{a0}(t)$ is the resonant frequency of the oscillator, $\alpha_{a}(t)$ corresponds to damping (loss), and $\omega_{ap}(t)$ is the plasma frequency ($a=\text{e,m}$). Notice that the resonator is driven by the average fields at the surface \eqref{Eq:AverageFields} as in the constitutive relation \eqref{Eq:ImpulseResponseConvTime}. We could solve the differential equation to obtain the impulse response for use in \eqref{Eq:ImpulseResponseConvTime} \cite{Zadeh:1950aa}. However, we will use \eqref{Eq:Lorentzian} directly, which takes into account the temporally dispersive nature of the surface~\cite{Caloz:2020aa}.  

All six parameters in \eqref{Eq:Lorentzian} are time-variant in general for a time-varying surface. Since the Lorentizan parameters are periodic in space and time, they can be written as Fourier series:
\begin{subequations}
	\begin{align}
		\omega_{a0}^2(x,t) &= \sum_{r=-\infty}^\infty\sum_{s=-\infty}^\infty \omega_{a0,rs} e^{j(n\omega_p t-m\beta_p x)}\label{Eq:LorentzianResonant}\\
	    \omega_{ap}^2(x,t) &= \sum_{r=-\infty}^\infty\sum_{s=-\infty}^\infty \omega_{ap,rs} e^{j(n\omega_p t-m\beta_p x)}\label{Eq:LorentzianPlasma}\\
	    \alpha_{a}(x,t) &= \sum_{r=-\infty}^\infty\sum_{s=-\infty}^\infty \alpha_{a,rs} e^{j(n\omega_p t-m\beta_p x)}\label{Eq:LorentzianAlpha}
	\end{align}\label{Eq:LorentzianParameters}
\end{subequations}

\noindent where $a=\text{(e,m)}$ for the electric and magnetic parameters, $\omega_p=2\pi/T$ is the temporal ``pumping frequency'' of the modulation, and $\beta_p=2\pi/p$ is the spatial frequency of the modulation. Note that in general, several resonators governed by \eqref{Eq:Lorentzian} may be required to have an accurate model of the metasurface, in which case responses of each of the resonators can be summed following the superposition principle, as the system is linear (i.e. $M_x=M_{x1}+M_{x2}+\cdots$ where each $M_{xn}$ is due to a resonator with unique parameters given by \eqref{Eq:LorentzianParameters}). We also note that while we consider TE fields in our analysis to demonstrate the method, for conciseness and simplicity, it can be straightforwardly extended to TM fields as well.
\section{Bloch-Floquet Expansion of Fields}\label{Sec:Formulation}

When a metasurface is periodic, the fields also become periodic, following Floquet's theorem. By expanding the fields in terms of space (and time) harmonics, we can produce a matrix equation to solve for the fields.

\setcounter{equation}{16}
\begin{figure*}[!b]
\noindent\rule{\textwidth}{0.5pt}%
\begin{subequations}%
	\begin{equation}%
		\begin{split}
	    	\sum_{m,n}\sum_{r,s}\left[(\omega_{e0,rs} - \delta(rs)\omega_{n-s}^2 + j\omega_{n-s} \alpha_{e,rs})Q_{m-r,n-s}-\frac{\omega_{ep,rs}}{2}(E_{t,m-r,n-s} + E_{r,m-r,n-s})\right]e^{j\Theta_{mn}}\\
	    		= \sum_{m,n} \omega_{ep,m-m_i,n-n_i}E_{i,m_in_i}e^{j\Theta_{mn}}\label{Eq:LorentzianQSub} %
		\end{split}
	\end{equation}
	\begin{equation}%
		\begin{split}%
			\sum_{m,n}\sum_{r,s}\left[(\omega_{m0,rs} - \delta(rs)\omega_{n-s}^2 + j\omega_{n-s} \alpha_{m,rs})M_{m-r,n-s}+\frac{\omega_{mp,rs}\cos\theta_{m-r,n-s}}{2\eta_0\cos\theta_{m_in_i}}(E_{t,m-r,n-s} - E_{r,m-r,n-s})\right]e^{j\Theta_{mn}} \\%
    		= - \sum_{m,n} \frac{\omega_{mp,m-m_i,n-n_i}}{\eta_0}E_{i,m_in_i}e^{j\Theta_{mn}}%
		\end{split}\label{Eq:LorentzianMSub}%
	\end{equation}%
	\label{Eq:EqnExpansions}%
\end{subequations}\\[-0.7cm]%
\begin{minipage}{\textwidth}%
\begin{minipage}[b][1cm][b]{0.48\textwidth}%
	\begin{subequations}[resume]%
 		\begin{equation}%
			\sum_{m,n} \left(j\mu_0\omega_nM_{mn} + E_{r0,mn}-E_{t0,mn}\right) e^{j\Theta_{mn}} = 0 \label{Eq:GSTCMSum}%
		\end{equation}%
	\end{subequations}%
\end{minipage}%
\begin{minipage}[b][1cm][b]{0.525\textwidth}%
	\begin{subequations}[resume]%
		\begin{equation}%
			\sum_{m,n} \left[\frac{j\omega_n}{c}Q_{mn} + (E_{r0,mn}+E_{t0,mn})\cos\theta_{mn}\right] e^{j\Theta_{mn}} = 0 \label{Eq:GSTCQSum}%
		\end{equation}%
	\end{subequations}%
\end{minipage}%
\end{minipage}\\
\end{figure*}
\setcounter{equation}{9}

\subsection{Expansion of Fields}\label{Sec:Expansion}


Applying Floquet's theorem, the electric fields can be expressed as a sum of space-time harmonics,
\begin{align}
	\Ev_a(x,z,t) 
		&= \mathbf{\hat{y}} \sum_{m=-\infty}^{\infty} \sum_{n=-\infty}^\infty E_{a0,mn} e^{j\Theta_{mn}} e^{\pm jk_{z,mn} z},\label{Eq:EFieldExpansion}
\end{align}
where $a={i,r,t}$ for the incident, reflected, and transmitted fields\footnote{The sign on $\pm jk_{z,mn} z$ is ($-$) for incident and transmitted harmonics and ($+$) for reflected harmonics.}, respectively, and where $\Theta_{mn}=(\omega_n t - k_{x,mn}x)$. Only a single harmonic is present for the incident field, $(m,n)=(m_i,n_i)$, corresponding to a plane wave.
Floquet's theorem prescribes that the transverse part of the wavevector ($k_x$) takes on discrete values determined by spatial periodicity, and the normal component ($k_z$) then follows from having a total magnitude $k_n$:
\begin{subequations}
	\begin{align}
		k_{x,m}=k_{n}\sin\theta_{mn} = k_0\sin\theta_{i} + m\beta_p,\label{Eq:kx}\\
		k_{z,mn}=k_{n}\cos\theta_{mn} = \sqrt{k_n^2-k_{x,mn}^2},\label{Eq:kz}
	\end{align}
\end{subequations}

This idea can be extended for the time harmonics, where the frequency can also only change by multiples of the pumping frequency, and the wavenumber changes accordingly:
\begin{gather}
	\omega_n = \omega_0 + n\omega_p,\label{Eq:wn}\\
	k_n = k_0 + n\frac{\omega_p}{c_0}=k_0\left(1+n\frac{\omega_p}{\omega_0}\right).\label{Eq:kn}
\end{gather}

Note that this allows for harmonics with negative frequencies, as well as potentially a dc ``harmonic'' if $\omega_0$ is an integer multiple of $\omega_p$. The harmonics with $\omega_n<0$ are not a cause for alarm, as $k_n<0$ for these harmonics and so the direction of propagation is physical (e.g. still $+z$ in the transmission region). The special case of $\omega_n=0$ is more questionable, but in the results, we show that this peculiarity also poses no problem.

\begin{figure}[h]
	\begin{overpic}[grid=false,trim={0cm 0 0cm 0},clip]{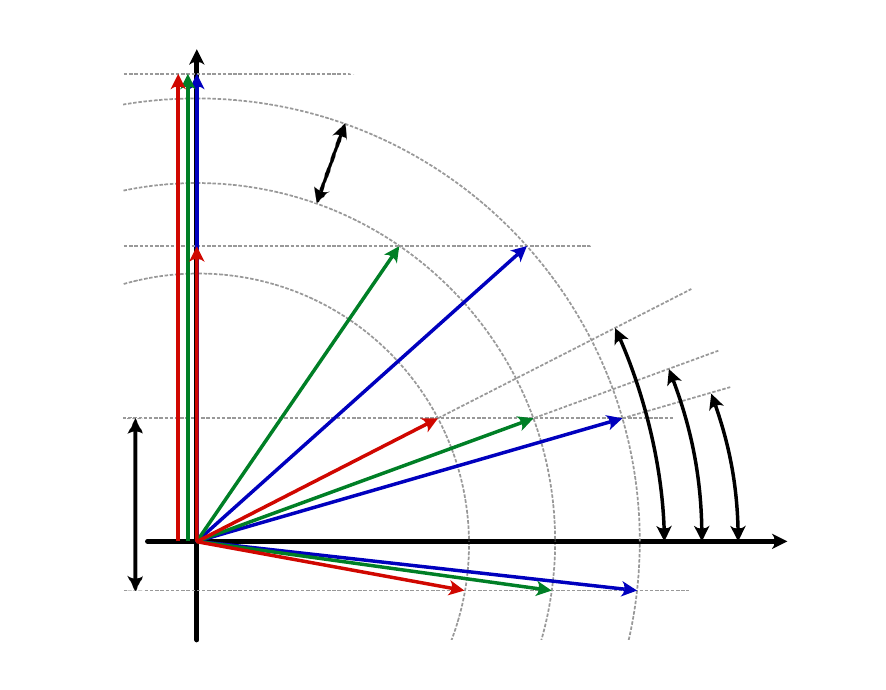}
		\put(11,20){\footnotesize \htext{$\beta_p$}}
		\put(41,57){\htext{\footnotesize $\frac{\omega_p}{c_0}$}}
		\put(90,16.2){\footnotesize $|\text{Re}\{k_z\}|$}
		\put(22.5,74){\footnotesize \htext{$k_x$}}
		\put(70,44){\tiny \htext{$\theta_{0,-1}$}}
		\put(75,38){\tiny $\theta_{i}$}
		\put(82,32.5){\tiny $\theta_{0,1}$}
		\put(49,32){\htext{\tiny $(0,-1)$}}
		\put(59,32){\htext{\tiny $(0,0)$}}
		\put(67,32){\htext{\tiny $(0,1)$}}
		\put(27,52){\htext{\tiny $(1,-1)$}}
		\put(47,52){\htext{\tiny $(1,0)$}}
		\put(60,52){\htext{\tiny $(1,1)$}}
		\put(11,70){\htext{\tiny $(2,-1)$}}
		\put(18,71){\htext{\tiny $(2,0)$}}
		\put(23,70){\tiny $(2,1)$}
		\put(50,8){\htext{\tiny $(-1,-1)$}}
		\put(60,8){\htext{\tiny $(-1,0)$}}
		\put(70,8){\htext{\tiny $(-1,1)$}}
		\put(11,45){\htext{\tiny $k_{-1}$}}
		\put(11,55){\htext{\tiny $k_0$}}
		\put(11,65.5){\htext{\tiny $k_{1}$}}
		\put(10,11){\htext{\tiny $k_{x,-1}$}}
		\put(10,30.5){\htext{\tiny $k_{x,0}$}}
		\put(11,50){\htext{\tiny $k_{x,1}$}}
		\put(41,69){\tiny $k_{x,2}$}
	\end{overpic}
	\caption{Space-time modulation produces scattered harmonics $(m,n)$ that exist at discrete frequencies $\omega_n$ (hence the $k_n$ circles) and discrete transverse spatial frequencies $k_{x,{m}}$ (horizontal lines). At each $k_{x,m}$, there are an infinite number of time harmonics with different $k_n$ and thus different angles of scattering $\theta_{mn}$.}
	\label{Fig:Harmonics}
\end{figure}

The angle of scattering for the harmonics are found by substituting \eqref{Eq:kn} into \eqref{Eq:kx}, which yields
\begin{align}
	\sin\theta_{mn} =\left( \frac{\sin\theta_{i}+m\beta_p/k_0}{1+n\omega_p/\omega_0}\right).\label{Eq:thetamn}
\end{align}
where $\theta_i=\theta_{00}$ is the angle of incidence. Each harmonic $(m,n)$ represents either an oblique propagating plane wave ($k_z\in\mathbb{R}$) or a surface wave ($k_z\in\mathbb{I}$), as illustrated in Figure~\ref{Fig:Harmonics}, where the real parts of the corresponding wavevectors are plotted. Using \eqref{Eq:kn}, we plot circles with constant $k_n$ (and $\omega_n$), while \eqref{Eq:kx} yields the horizontal lines that represent the allowed values $k_{x,m}$ (which is always purely real). The intersection of the circles and lines represents possible \textit{propagating} space-time harmonics. \textit{Surface waves} on the other hand lie on the vertical axis and not necessarily on a $k_n$ circle.

Finally, the magnetic field can also be expanded,
\begin{align}
\mathbf{H}_{a}(x,z,t) 
	&= \frac{1}{\eta_0}\sum_{m=-\infty}^{\infty} \sum_{n=-\infty}^\infty
	\left[\begin{array}{@{}l@{}}
		(\sin\theta_{mn}\mathbf{\hat{z}} \pm \cos\theta_{mn}\mathbf{\hat{x}})\\
		~\times E_{a0,mn} e^{j\Theta_{mn}} e^{\pm jk_{z,mn} z}
	\end{array}\right],
	\label{Eq:HFieldExpansion}
\end{align}
along with the polarization densities, which are
\begin{subequations}\label{Eq:PolarizationExpansion}
	\begin{align}
		Q_y(x,t) = \sum_{m=-\infty}^{\infty} \sum_{n=-\infty}^\infty Q_{mn}e^{j\Theta_{mn}},\label{Eq:QExpansion}
	\end{align}
	\begin{align}
		M_x(x,t) = \sum_{m=-\infty}^{\infty} \sum_{n=-\infty}^\infty M_{mn}e^{j\Theta_{mn}}.\label{Eq:MExpansion}
	\end{align}
\end{subequations}

\subsection{Matrix Formulation}\label{Sec:EquationFormulation}


We begin by substituting $\Ev_a(x,0,t)$ from \eqref{Eq:EFieldExpansion} into \eqref{Eq:AverageFields} for the incident and scattered fields, and the resulting average field into \eqref{Eq:LorentzianQ}. Similarly, the expansion of $Q_y(x,t)$ from \eqref{Eq:QExpansion} is substituted into \eqref{Eq:LorentzianQ}, providing a set of infinite equations \eqref{Eq:LorentzianQSub}. This procedure is repeated with the magnetic field and \eqref{Eq:LorentzianM}, producing \eqref{Eq:LorentzianMSub}.


Next, the expansions of the fields and polarization densities are also substituted into the GSTC equations \eqref{Eq:GSTCScalar}, producing \eqref{Eq:GSTCMSum} and \eqref{Eq:GSTCQSum}. This leaves us with four sets of infinite equations for the four sets of harmonics, $E_{r,mn}$, $E_{t,mn}$, $Q_{mn}$, and $M_{mn}$. To make the problem tractable, the harmonics can be truncated to $-M<m<M$ and $-N<n<N$, which corresponds to $(2M+1)$ space harmonics, each of which has $(2N+1)$ time harmonics. This truncation assumes that the selected number of harmonics is sufficient, and this assumption must be verified after computation in the form of a series convergence. Furthermore, the finite system of equations can be written in matrix form for implementation in code, as described in Appendix~\ref{Sec:AppendixFormulation}. This allows solving for the four sets of $(2N+1)\cdot(2M+1)$ unknown harmonics.

\subsection{Extension to an Arbitrary Incident Field}\label{Sec:FormulationGeneralExcitation}


The method presented in Section~\ref{Sec:EquationFormulation} allows solving the scattering due to a plane wave excitation at an angle $\theta_{i}$, but it can be extended to arbitrary excitations, such as a single Gaussian beam (spatial distribution) or a Gaussian pulse (temporal shape). Let us denote the arbitrary incident field as $E_\text{i,\text{tot}}(x,z,t)$. The metasurface responds to the field at $z=0$, where we can decompose $E_\text{i,\text{tot}}$ into plane waves, using a Fourier transform,
\setcounter{equation}{17}
\begin{align}
	E_{i}(p,q) = \frac{1}{(2\pi)^2} \int_{-\infty}^\infty\int_{-\infty}^\infty E_{i,\text{tot}}(x,0,t)e^{-j(p\omega_s t-qk_{xs} x)}\,dx\,d\omega,\label{Eq:ArbitraryDecomposition}
\end{align}
\noindent which is densely sampled (small $\omega_s$ and $k_{x,s}$) to yield a good approximation of the finite signal. Following this, \eqref{Eq:EqnExpansions} can be solved for each of the plane waves with $\omega_0=p\omega_s$ and $\theta_i=\sin^{-1}(qk_{xs}c_0/\omega_0)$, yielding $\Ev_{a,pq}(x,z,t)$. These are simply summed to produce the scattered fields:
\begin{align}
	\Ev_{a,\text{tot}}(x,z,t) \approx \sum_{p,q} E_{i}(p,q) \Ev_{a,pq}(x,z,t)
\end{align}

\noindent where $a=(t,r)$. Of course, the 2D Fourier transform \eqref{Eq:ArbitraryDecomposition} can be simplified to a 1D transform if the input signal is monochromatic or spatially uniform.

\section{Results}\label{Sec:Results}


To demonstrate the proposed method, we consider three cases of periodically modulated surfaces: space-only modulation ($\omega_p=0$), time-only modulation ($\beta_p=0$), and general space-time modulation.

\subsection{Space-Only Modulation}\label{Sec:PureSpace}

\begin{figure}[h!]
	\centering
	\subfloat[Total field, Floquet]{%
		\begin{overpic}[width=0.5\columnwidth,grid=false,trim={1.3cm -0.3cm 0.5cm 0},clip]{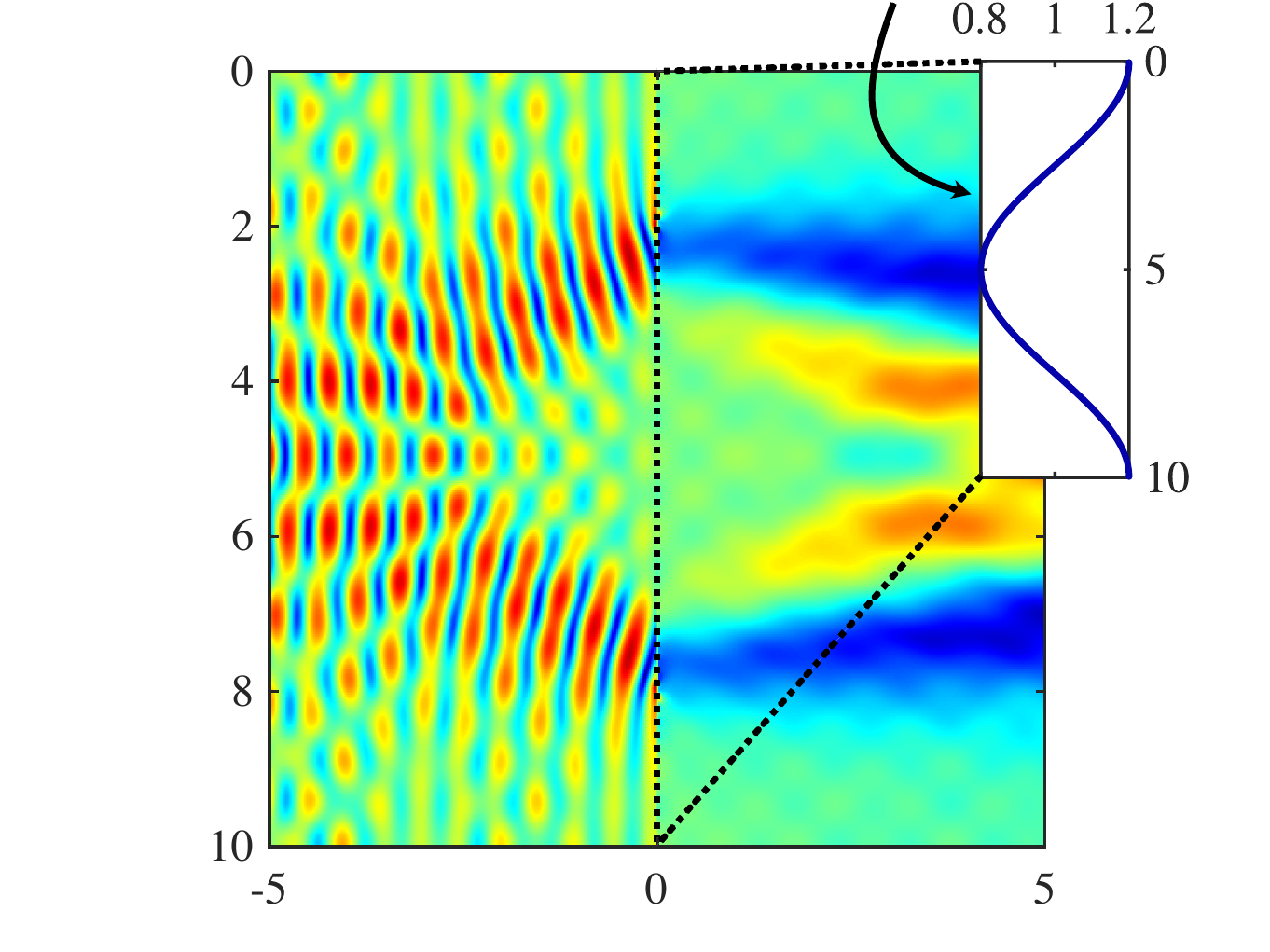}
			\put(49,0){\htext{\footnotesize $z/\lambda$}}
			\put(5,45){\vtext{\footnotesize $x/\lambda$}}
			\put(85,92){\htext{\footnotesize $\omega_{e0,m0}(x)/\omega_{e0,m0}$}}
			\put(98,62){\vtext{\tiny $x/\lambda$}}
		\end{overpic}
		\label{Fig:SpatialModCosineFlo}
 	}
 	\subfloat[Total field, FDFD]{%
 		\begin{overpic}[width=0.5\columnwidth,grid=false,trim={1.0cm -0.3cm 0.8cm 0},clip]{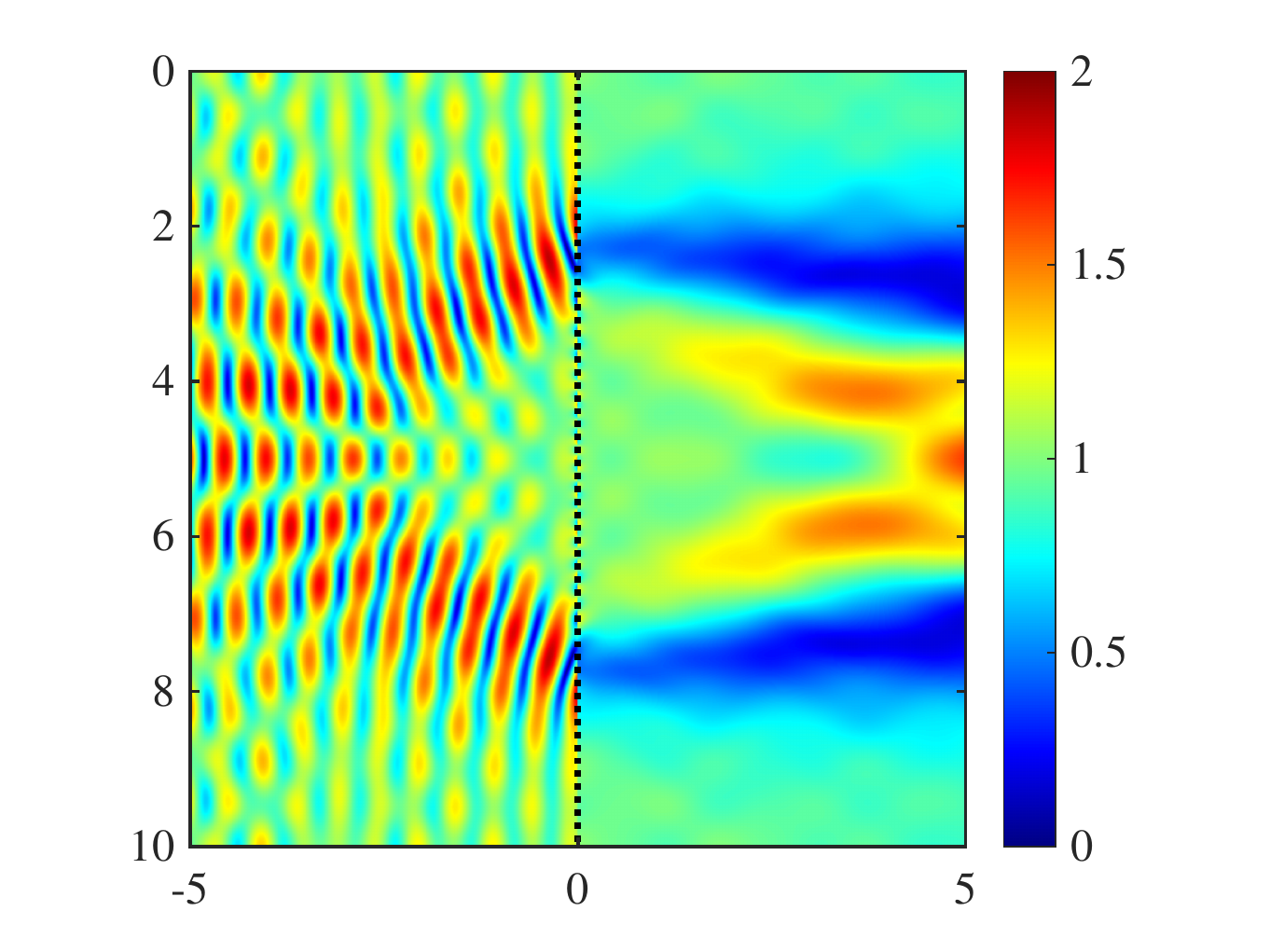}
			\put(45,0){\htext{\footnotesize $z/\lambda$}}
			\put(2,44){\vtext{\footnotesize $x/\lambda$}}
			\put(95,46){\vtext{\tiny $|E|/|E_{i,00}|$}}
		\end{overpic}
		\label{Fig:SpatialModCosineFDFD}
  	}\\
  	\subfloat[DFT of total field]{%
 		%
 		\begin{overpic}[width=0.5\columnwidth,grid=false,trim={0.2cm -0.3cm 0.6cm 0},clip]{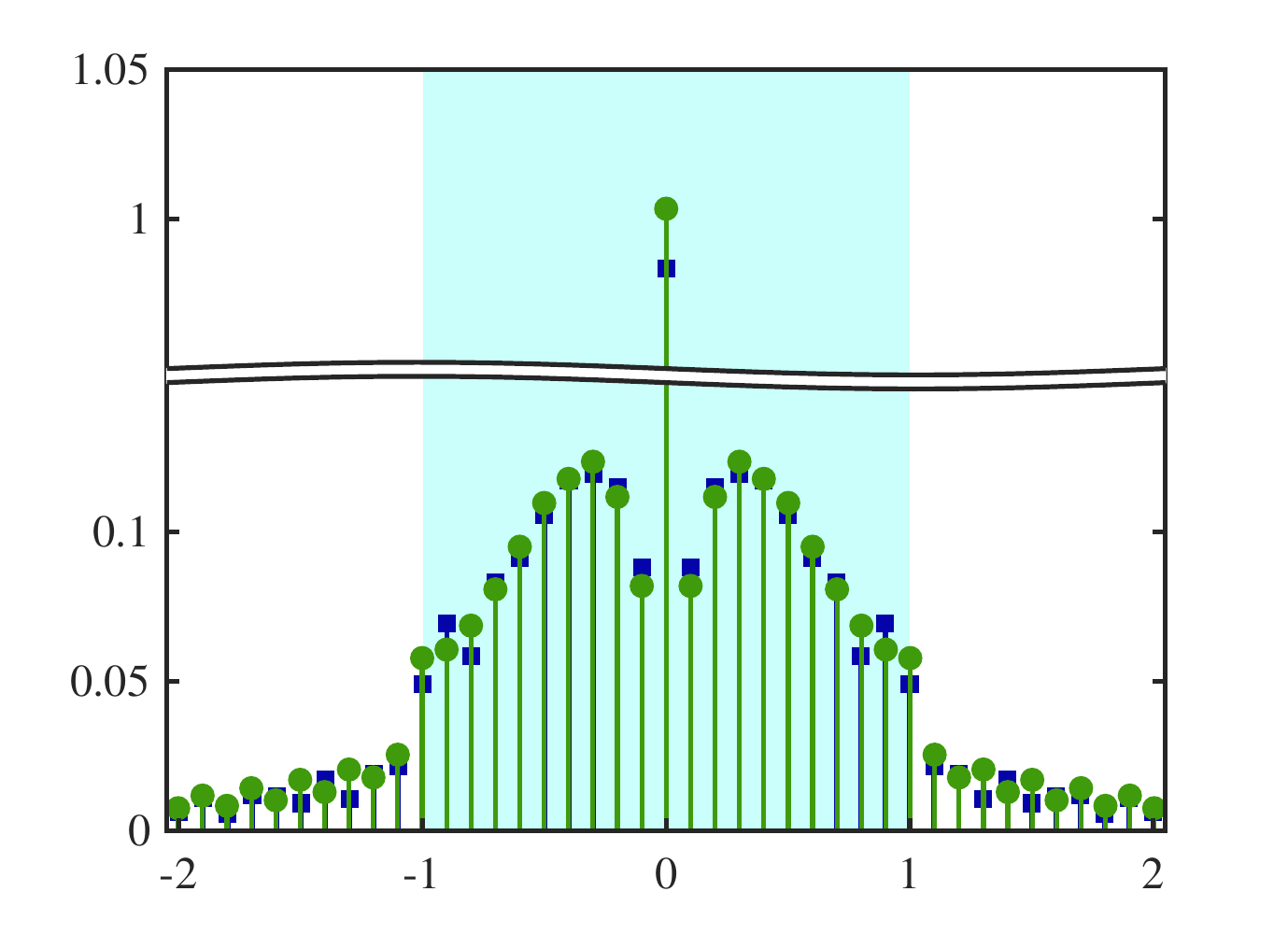}
			\put(54,0){\htext{\footnotesize $k_x/k_0$}}
			\put(0,42){\vtext{\footnotesize $|E|$}}
			\put(52,76.5){\htext{\small $z=-\lambda/10$}}
		\end{overpic}
 		\begin{overpic}[width=0.5\columnwidth,grid=false,trim={0.2cm -0.3cm 0.6cm 0},clip]{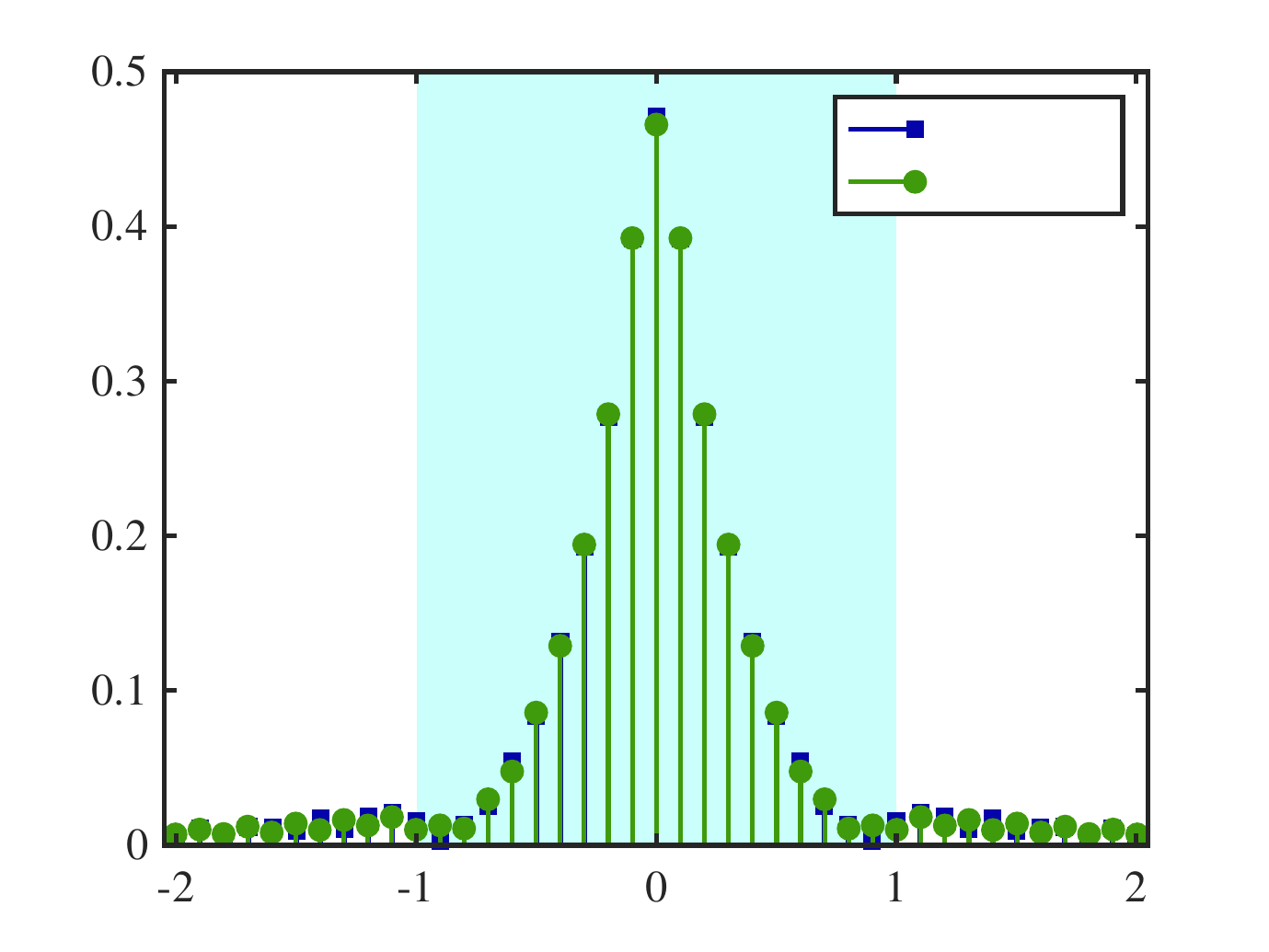}
			\put(54,0){\htext{\footnotesize $k_x/k_0$}}
			\put(0,42){\vtext{\footnotesize $|E|$}}
			\put(52,78){\htext{\small $z=+\lambda/10$}}
			\put(77,65.5){\tiny FDFD}
			\put(77,69.5){\tiny Floquet}
		\end{overpic}
	 	\label{Fig:SpatialModCosineHarmonics}
  	}
	\caption{A plane wave with $\omega_0=2\pi(\SI{230}{THz})$ is normally incident on a metasurface where the resonant frequencies are spatially modulated as $\omega_{e0,m0}(x)=\omega_{e0,m0}[1+\Delta_{e,m}\cos(\beta_px){]}$. The total field $|E|$ for one period $0<x<p$ is shown in (a) and (b), while the amplitudes of the spatial harmonics are shown in (c). We use $M=100$ for the Floquet solution, while the FDFD was performed at 100 divisions per wavelength. The parameters of the modulation are $\Delta_{e,m}=0.2$ and $\beta_p=k_0/10$, and the Lorentzian susceptibilities have nominal values $\omega_{e0}=2\pi(\SI{230}{THz})$, $\omega_{m0}=2\pi(\SI{215}{THz})$, $\omega_{ep}=\omega_{mp}=\SI{301e9}{rad/s}$ and $\alpha_e=\alpha_m=\SI{7.54e12}{s^{-1}}$.}
	\label{Fig:SpatialModCosine}
\end{figure}
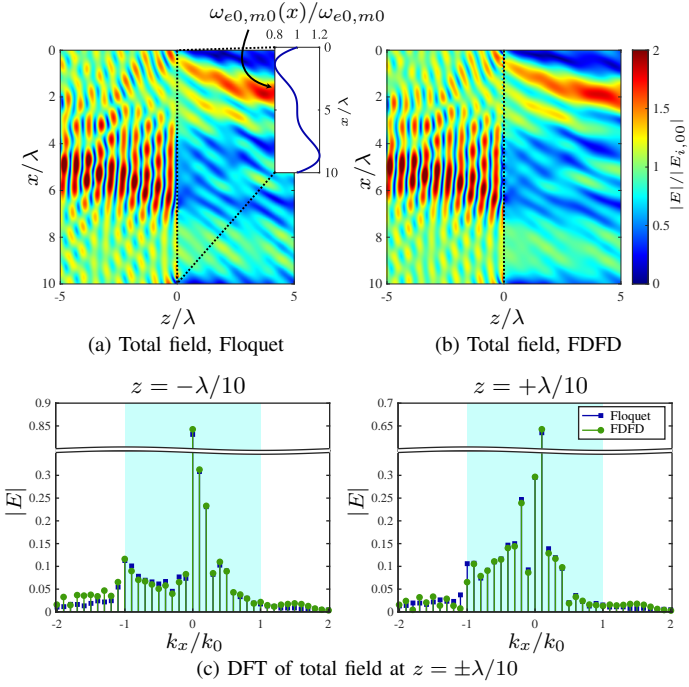
\begin{figure}[h]
	\centering
	\subfloat[Total field, Floquet]{%
		\begin{overpic}[width=0.5\columnwidth,grid=false,trim={1.3cm -0.3cm 0.5cm 0},clip]{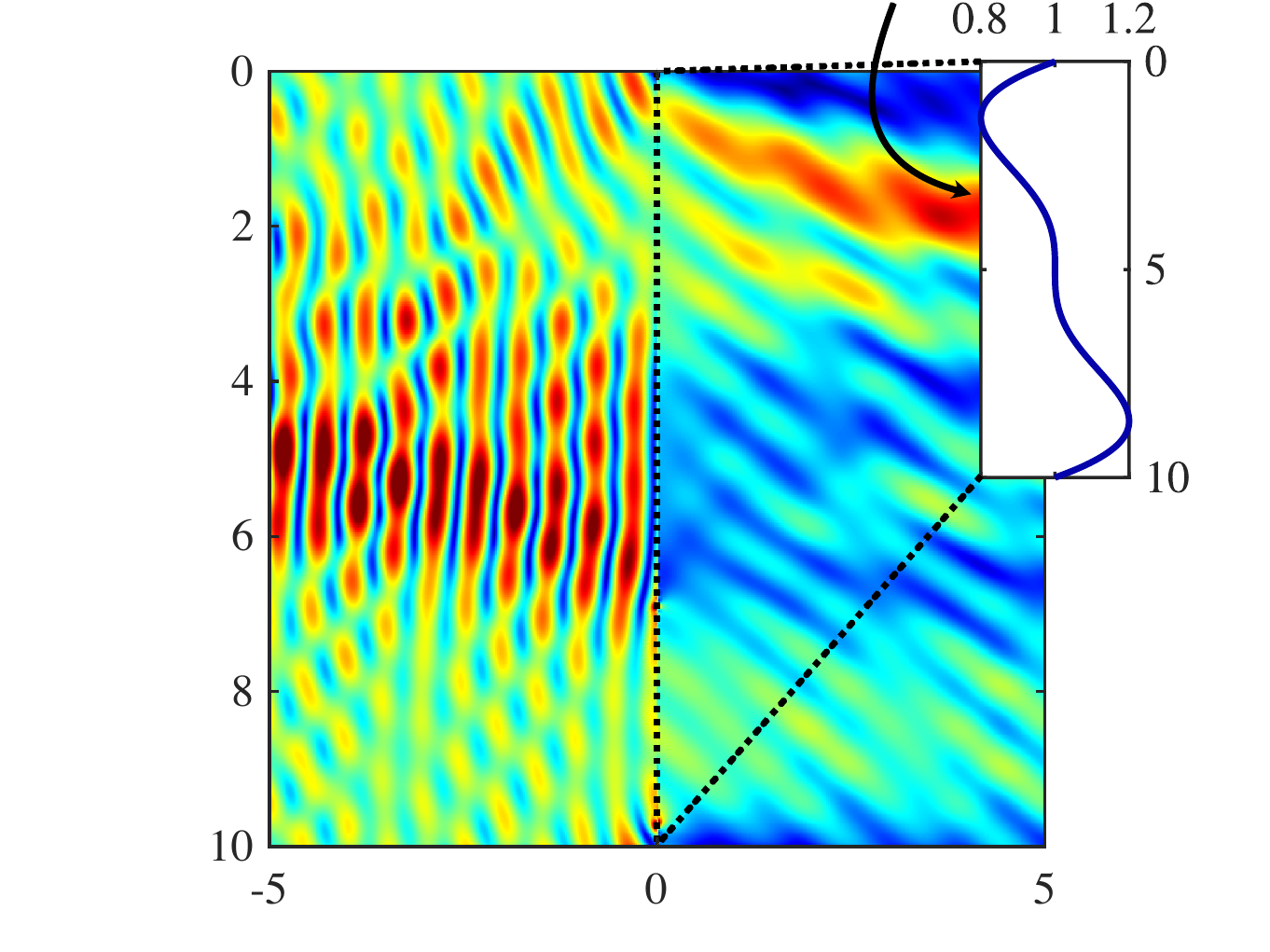}
			\put(49,0){\htext{\footnotesize $z/\lambda$}}
			\put(5,45){\vtext{\footnotesize $x/\lambda$}}
			\put(85,92){\htext{\footnotesize $\omega_{e0,m0}(x)/\omega_{e0,m0}$}}
			\put(98,62){\vtext{\tiny $x/\lambda$}}
		\end{overpic}
 	}
 	\subfloat[Total field, FDFD]{%
 		\begin{overpic}[width=0.5\columnwidth,grid=false,trim={1.0cm -0.3cm 0.8cm 0},clip]{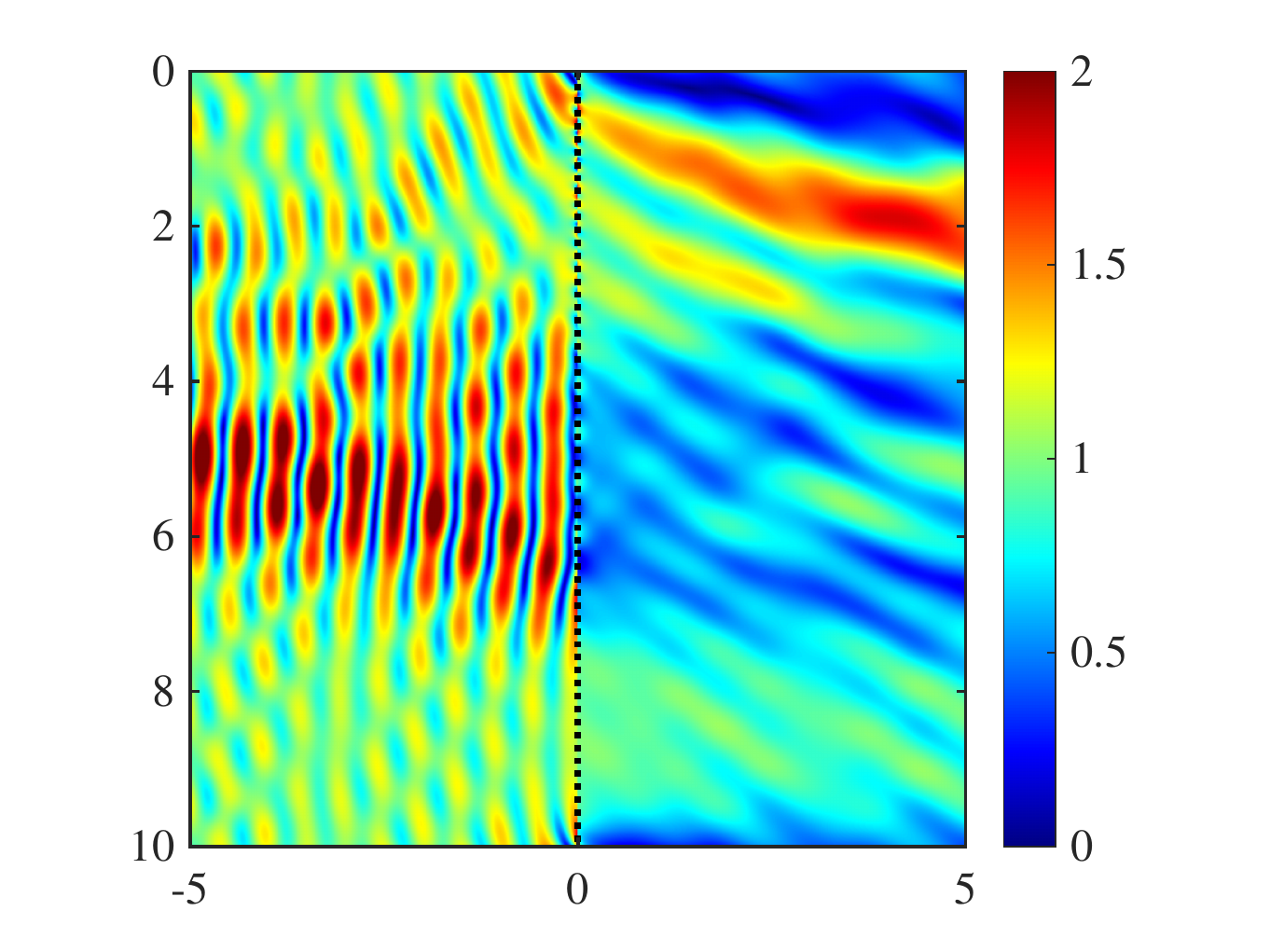}
			\put(45,0){\htext{\footnotesize $z/\lambda$}}
			\put(2,44){\vtext{\footnotesize $x/\lambda$}}
			\put(95,46){\vtext{\tiny $|E|/|E_{i,00}|$}}
		\end{overpic}
  	}\\
  	\subfloat[DFT of total field at $z=\pm \lambda/10$]{%
  		\begin{overpic}[width=0.5\columnwidth,grid=false,trim={0.2cm -0.3cm 0.6cm 0},clip]{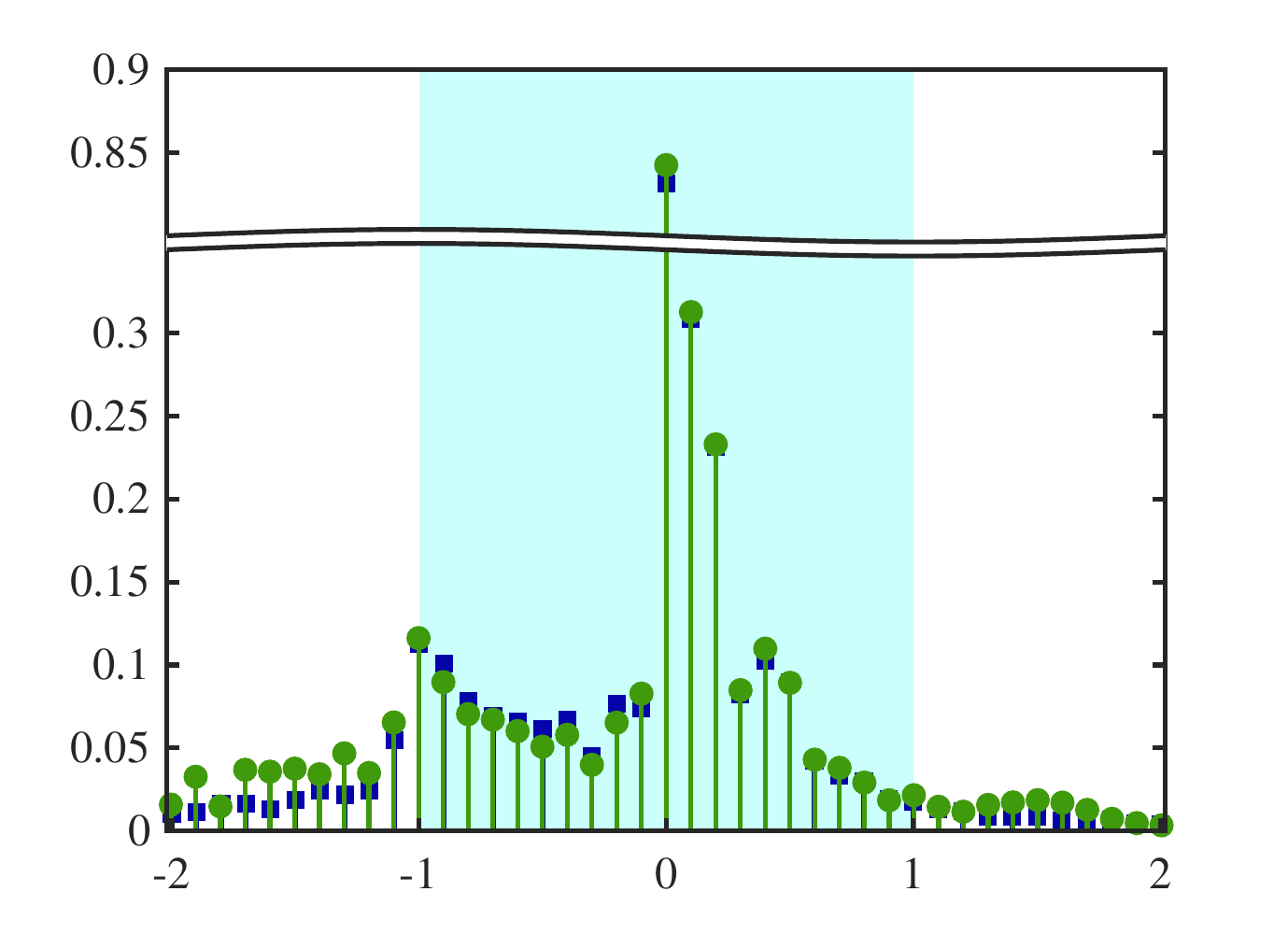}
			\put(54,0){\htext{\footnotesize $k_x/k_0$}}
			\put(0,42){\vtext{\footnotesize $|E|$}}
			\put(52,76.5){\htext{\small $z=-\lambda/10$}}
		\end{overpic}
 		\begin{overpic}[width=0.5\columnwidth,grid=false,trim={0.2cm -0.3cm 0.6cm 0},clip]{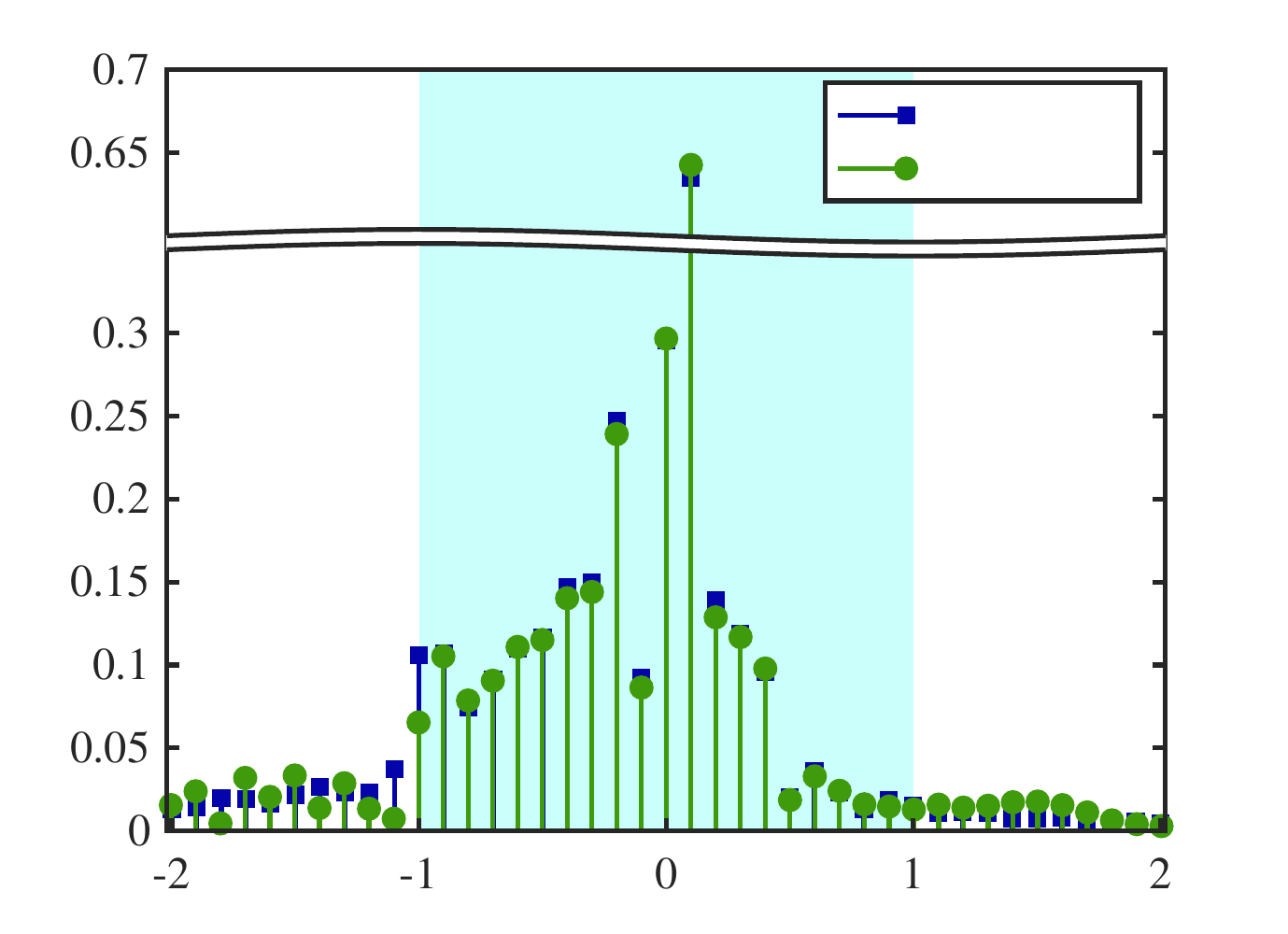}
			\put(54,0){\htext{\footnotesize $k_x/k_0$}}
			\put(0,42){\vtext{\footnotesize $|E|$}}
			\put(52,76.5){\htext{\small $z=+\lambda/10$}}
			\put(76,64.5){\tiny FDFD}
			\put(76,68.5){\tiny Floquet}
		\end{overpic}
  	}
	\caption{Repeating the simulations from Fig.~\ref{Fig:SpatialModCosine} with the modulation changed to $\omega_{e0,m0}(x)=\omega_{e0,m0}[1-\frac{5}{2}\Delta_{e,m}|\text{sawtooth}(\beta_px)|\sin(\beta_px){]}$, once again with $\Delta_{e,m}=0.2$ and $\beta_p=k_0/10$.\protect\footnotemark}
	\label{Fig:SpatialModSawtooth}
\end{figure}

First, we consider a spatial modulation of the electric and magnetic resonant frequencies, $\omega_{e0}$ and $\omega_{m0}$, using a cosine profile (see inset in Fig.~\ref{Fig:SpatialModCosineFlo}). The modulation harmonics are calculated using \eqref{Eq:LorentzianResonant} and the system of equations \eqref{Eq:EqnExpansions} are solved for a normally-incident plane wave with $M=100$ and $N=0$, for a total of 201 harmonics, 
producing the magnitude plotted in Fig.~\ref{Fig:SpatialModCosineFlo}.
Since the scattered fields are monochromatic, a frequency domain simulator can be used to verify the Floquet result; a finite-difference frequency-domain (FDFD) simulation was run, producing a field magnitude in agreement with the Floquet solution (Fig.~\ref{Fig:SpatialModCosineFDFD}). The spatial harmonics were also compared with a discrete Fourier transform (DFT) of both fields at $z=\pm \lambda/10$ on both sides of the surface, with good agreement for both propagating harmonics ($|k_x|<k_0$, highlighted in blue) and evanescent harmonics ($|k_x|>k_0$). Next, a more complex asymmetrical profile was used for the modulation. Using the same procedure, the fields and harmonics are plotted in Fig.~\ref{Fig:SpatialModSawtooth}. Even though $\theta_i=\SI{0}{\degree}$, the scattering primarily occurs towards $+x$ with $k_x\geq 0$ harmonics being dominant, which can be expected for this surface which imparts an asymmetric phase variation \cite{Ding:2017aa}.

\footnotetext{The sawtooth function from MATLAB was used with a period $2\pi$ and peaks $-1$ and $1$.}

\begin{figure}[h]
	\centering
	\subfloat[Total field, Floquet]{%
		\begin{overpic}[width=0.5\columnwidth,grid=false,trim={2.5cm 0 1.5cm 0.3cm},clip]{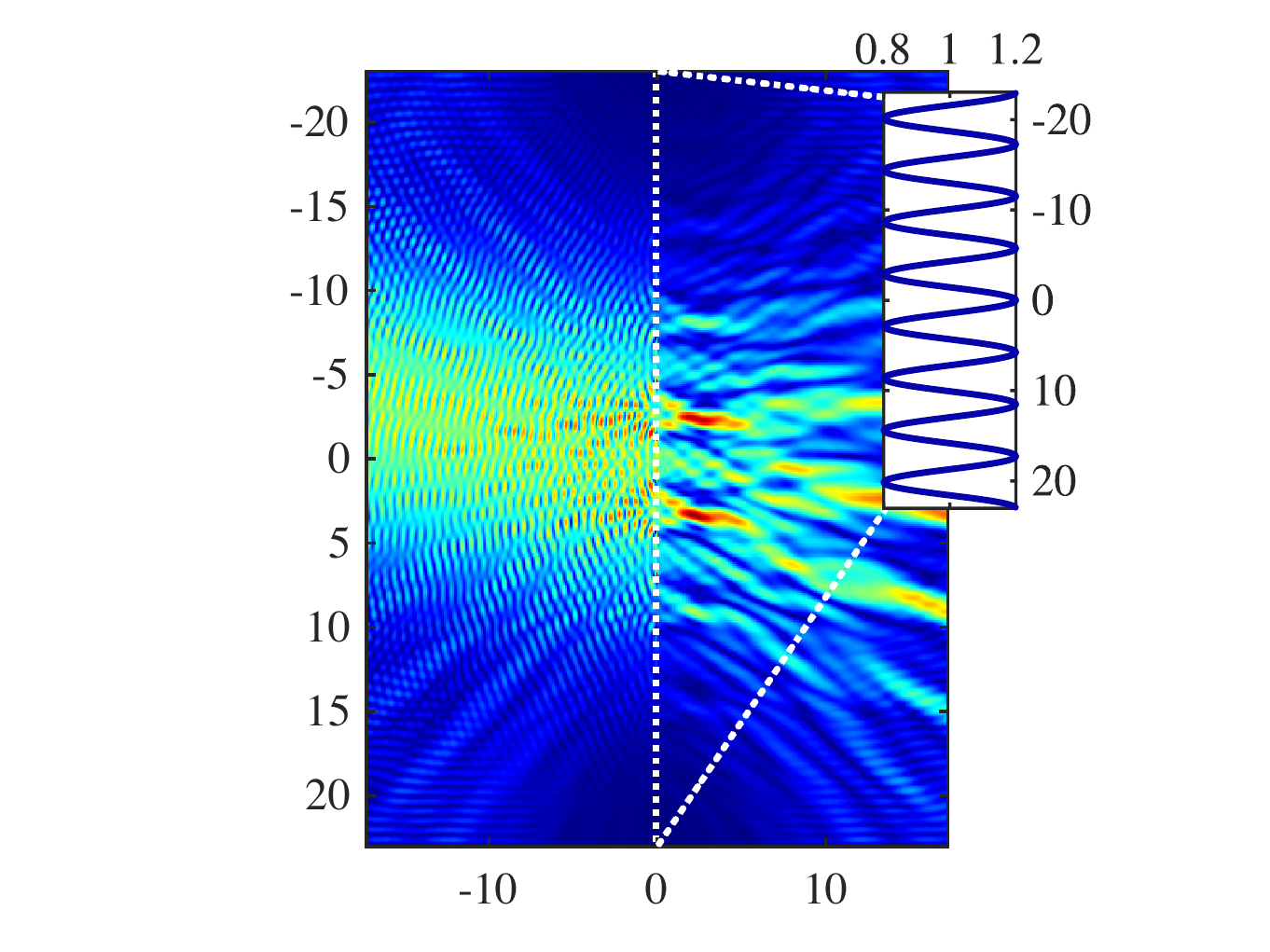}
			\put(46,0){\htext{\footnotesize $z/\lambda$}}
			\put(5,51){\vtext{\footnotesize $x/\lambda$}}
			\put(77,102){\htext{\tiny $\omega_{e0,m0}(x)/\omega_{e0,m0}$}}
			\put(92,68){\vtext{\tiny $x/\lambda$}}
		\end{overpic}
 	}
 	\subfloat[Total field, FDFD]{%
 		\begin{overpic}[width=0.5\columnwidth,grid=false,trim={2cm 0 2cm 0.3cm},clip]{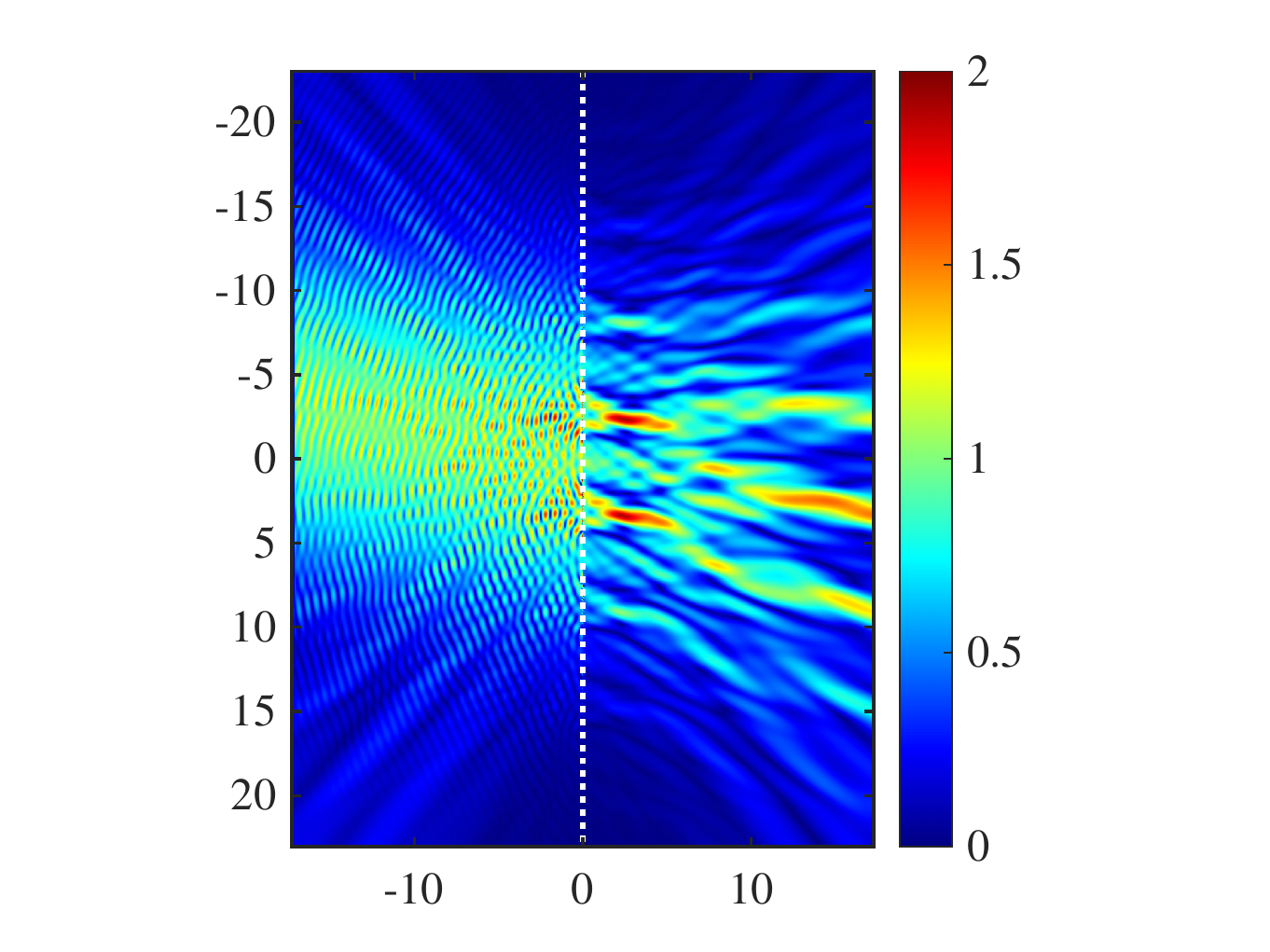}
			\put(42,0){\htext{\footnotesize $z/\lambda$}}
			\put(2,51){\vtext{\footnotesize $x/\lambda$}}
			\put(90,52){\vtext{\tiny $|E|/|E_{i,00}|$}}
		\end{overpic}
  	}\\
  	\subfloat[DFT of total field at $z=\pm \lambda/10$]{%
  		\begin{overpic}[width=0.5\columnwidth,grid=false,trim={0.2cm -0.3cm 0.6cm 0},clip]{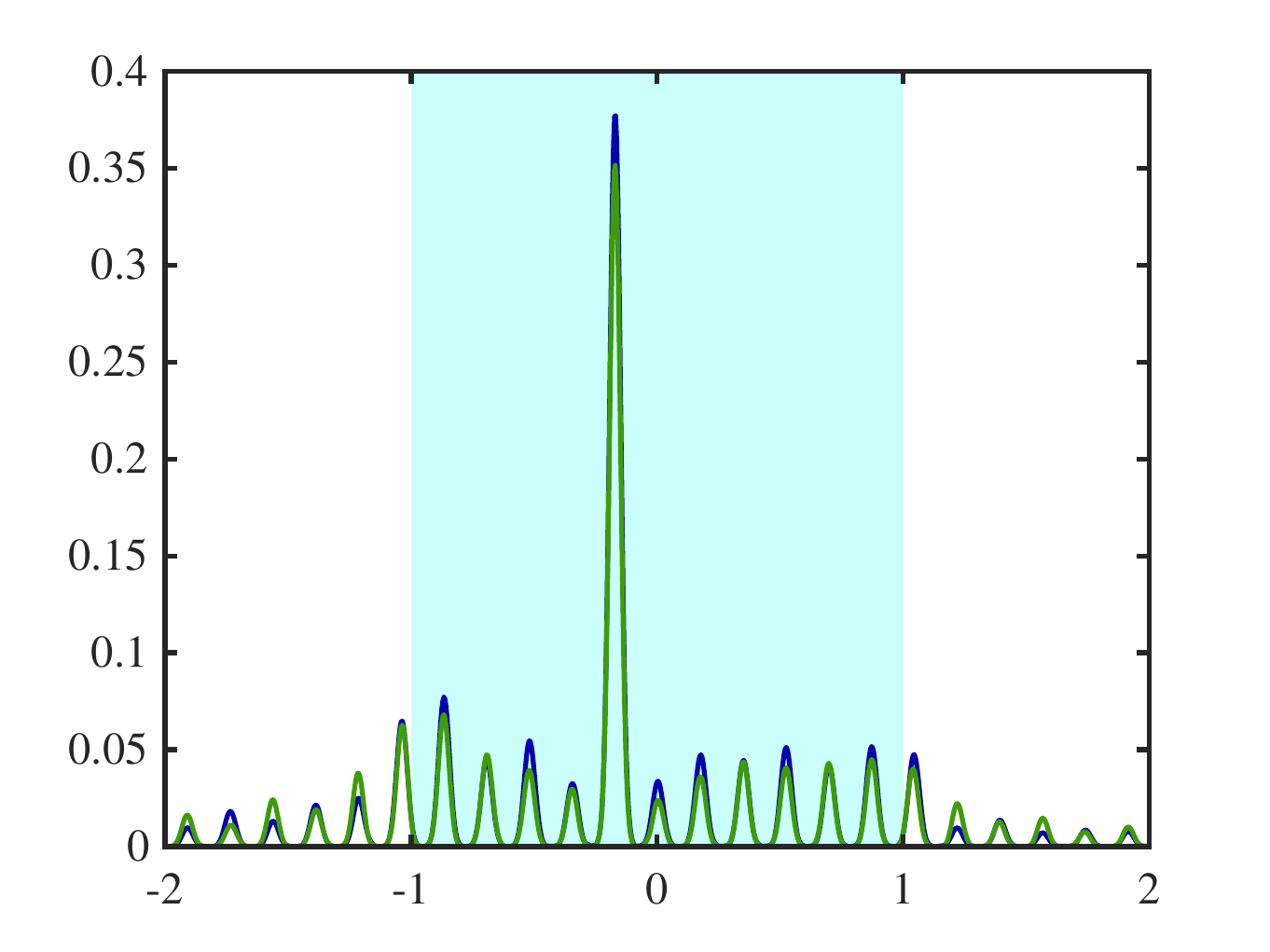}
			\put(54,0){\htext{\footnotesize $k_x/k_0$}}
			\put(0,44){\vtext{\footnotesize $|E|$}}
			\put(52,79){\htext{\small $z=-\lambda/10$}}
		\end{overpic}
 		\begin{overpic}[width=0.5\columnwidth,grid=false,trim={0.2cm -0.3cm 0.6cm 0},clip]{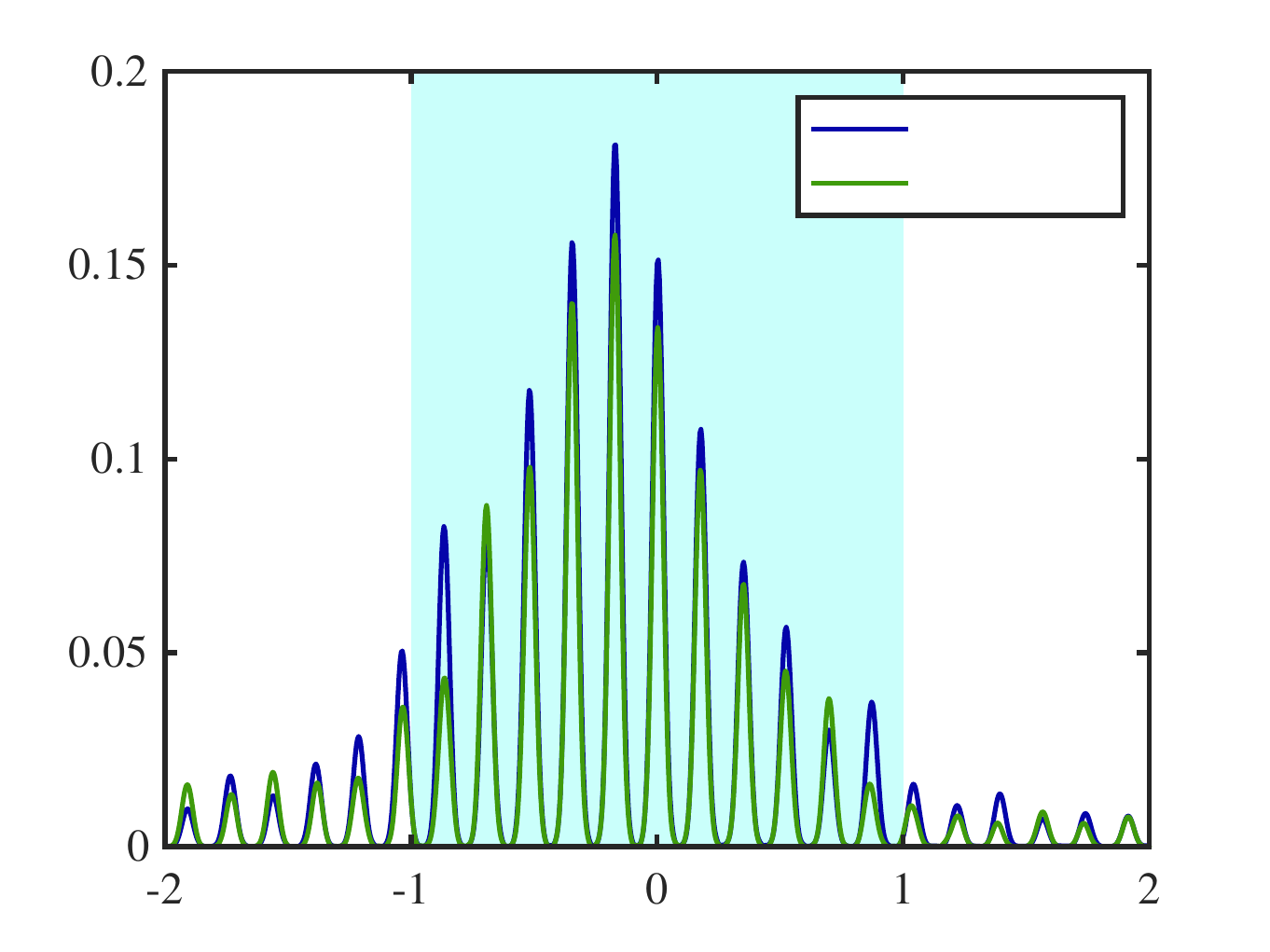}
			\put(54,0){\htext{\footnotesize $k_x/k_0$}}
			\put(0,44){\vtext{\footnotesize $|E|$}}
			\put(52,79){\htext{\small $z=+\lambda/10$}}
			\put(76,65.5){\tiny FDFD}
			\put(76,69.5){\tiny Floquet}
		\end{overpic}
  	}
	\caption{Shows the total field $|E|$ produced when a Gaussian beam with waist $10\lambda$ is incident at $\theta_i=\SI{-10}{\degree}$ onto the surface from Fig.~\ref{Fig:SpatialModCosine}. The modulation is $\omega_{e0,m0}(x)=\omega_{e0,m0}[1+\Delta_{e,m}\cos(\beta_px){]}$ with $\beta_p=k_0/5.76$ and $\Delta_{e,m}=0.2$.}
	\label{Fig:SpatialGaussian}
\end{figure}
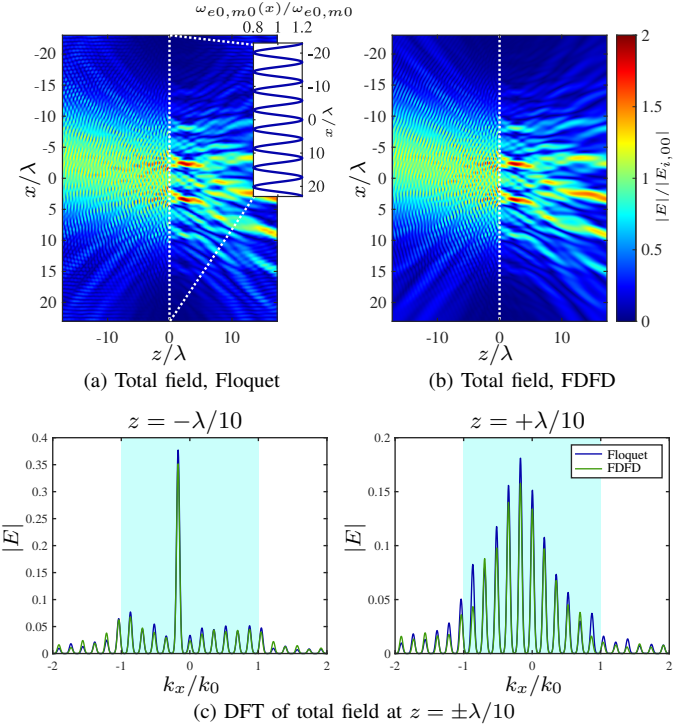

Finally, Fig.~\ref{Fig:SpatialGaussian} shows an example of a Gaussian beam incident on a surface with a cosine modulation profile. In this case, the angle of incidence is $\theta_i=\SI{-10}{\degree}$, and the surface is designed so that the $m=-1$ harmonic is scattered normally (we find $\beta_p=k_0/5.76$ from \eqref{Eq:thetamn}). We use a beam waist of $10\lambda$ and decompose the field using \eqref{Eq:ArbitraryDecomposition} into 23 plane waves (found to be sufficient in representing the spatial Gaussian profile). After computing the fields for each of these plane waves and summing the total fields, the Floquet method shows good agreement with the FDFD result. The slight discrepency between the two methods becomes smaller as the number of harmonics is increased for the Floquet method, and the FDFD mesh is made more dense. Note that while one harmonic is scattered towards $\theta_{-1,0}=\SI{0}{\degree}$, harmonics are also scattered in other directions. We will show in Section~\ref{Sec:ResultsSpaceTime} that if the spatial modulation is coupled with a time modulation, the harmonic at \SI{0}{\degree} can be converted to a different frequency to isolate it from the other spatial harmonics.

\subsection{Time-Only Modulation}

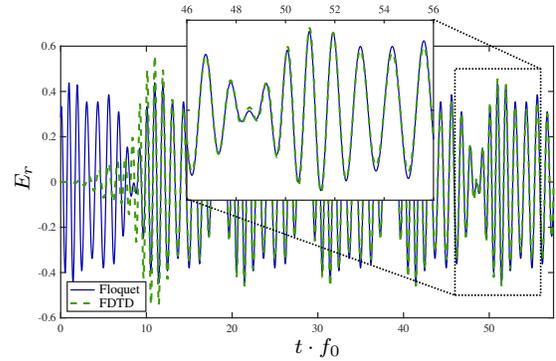
\begin{figure}
	\centering
		\begin{overpic}[width=0.4\textwidth,grid=false,trim={2cm 0.3cm 2cm 1cm},clip]{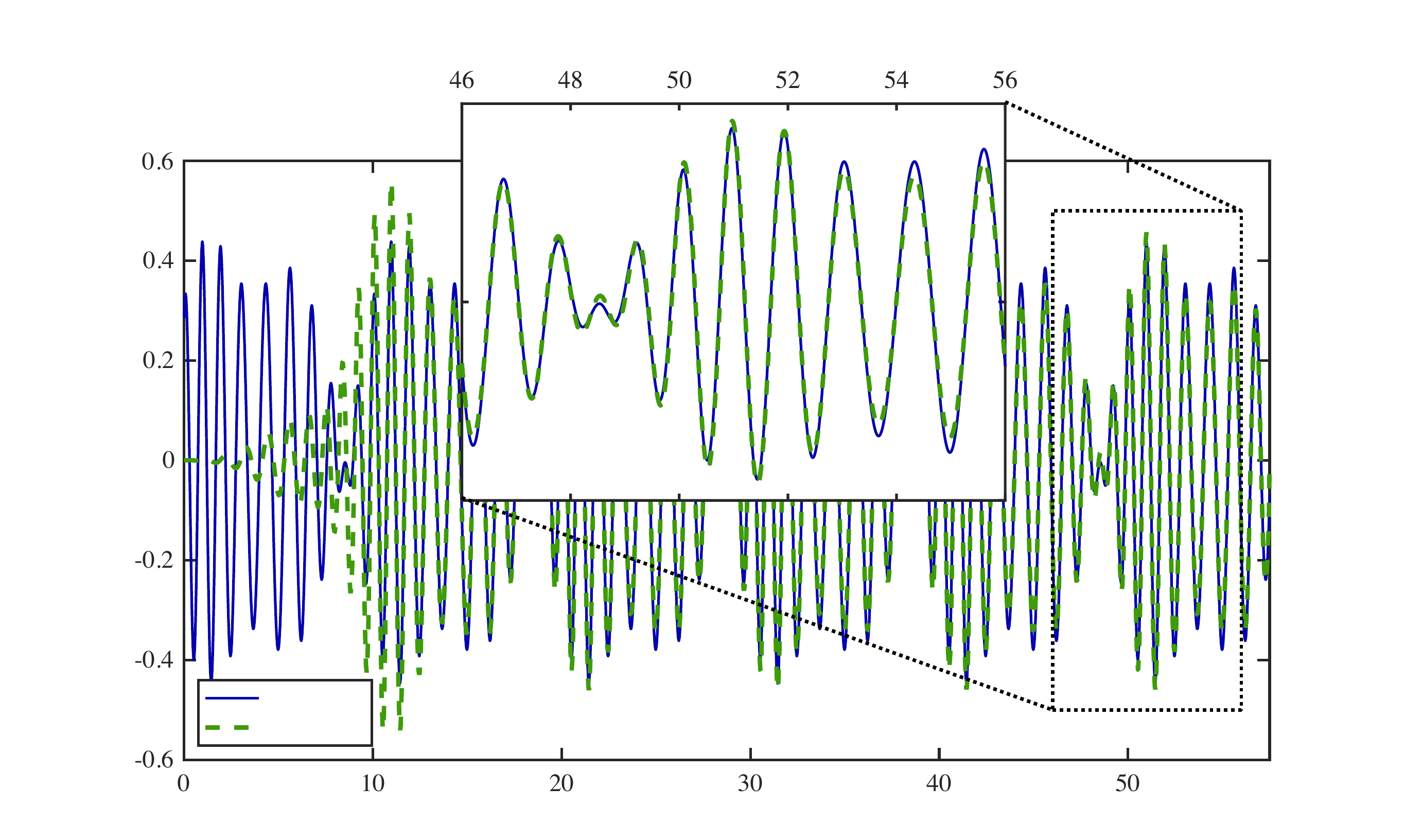}
			\put(54,-1){\htext{\footnotesize $t\cdot f_0$}}
			\put(0,30){\vtext{\footnotesize $E_r$}}
			\put(14,9.8){\tiny Floquet}
			\put(14,7.3){\tiny FDTD}
		\end{overpic}
	\caption{Shows the reflected field at the surface ($z=0^-$) for a uniform surface that is time-modulated. For validation, an FDTD simulation is run until it reaches a steady state (inset). The transmitted field is treated likewise. The resonant frequencies are modulated, $\omega_{e0,m0}(x)=\omega_{e0,m0}[1+\Delta_{e,m}\cos(\omega_pt){]}$, with $\omega_p=\omega_0/10$ and $\Delta_{e}=\Delta_m=0.2$.}
	\label{Fig:TimeSteadyState}
\end{figure}
\begin{figure}
	\centering
	\subfloat[$\omega_p=\omega_0/10$, $\Delta_{e}=\Delta_m=0.2$]{%
  		\begin{overpic}[width=0.25\textwidth,grid=false,trim={0cm 0cm 0.5cm 0cm},clip]{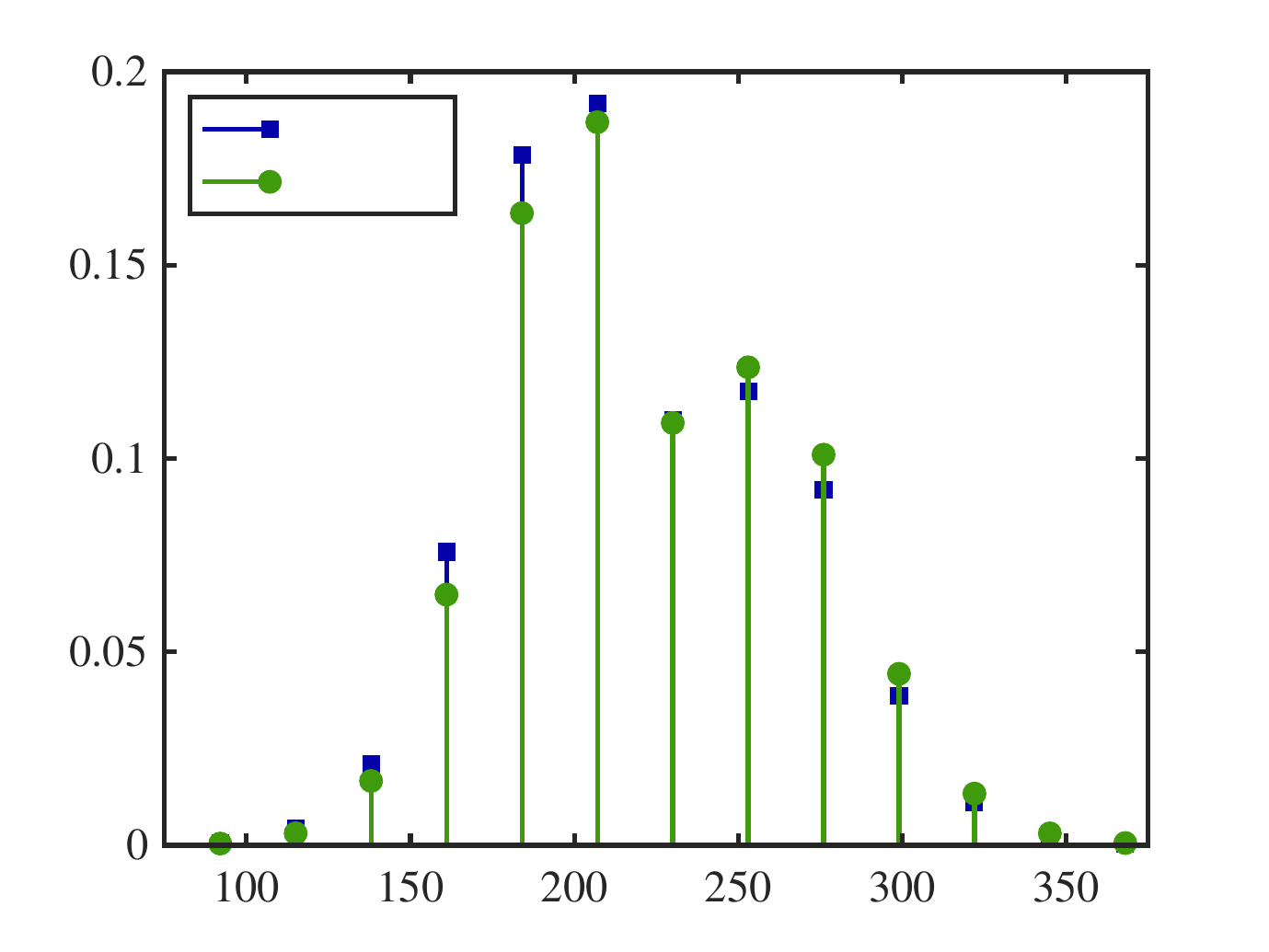}
			\put(52,-1){\htext{\footnotesize $f$ (THz)}}
			\put(3,40){\vtext{\footnotesize $|E_r|$}}
			\put(24,66.5){\tiny Floquet}
			\put(24,61.5){\tiny FDTD}
		\end{overpic}
		\begin{overpic}[width=0.25\textwidth,grid=false,trim={0cm 0cm 0.5cm 0cm},clip]{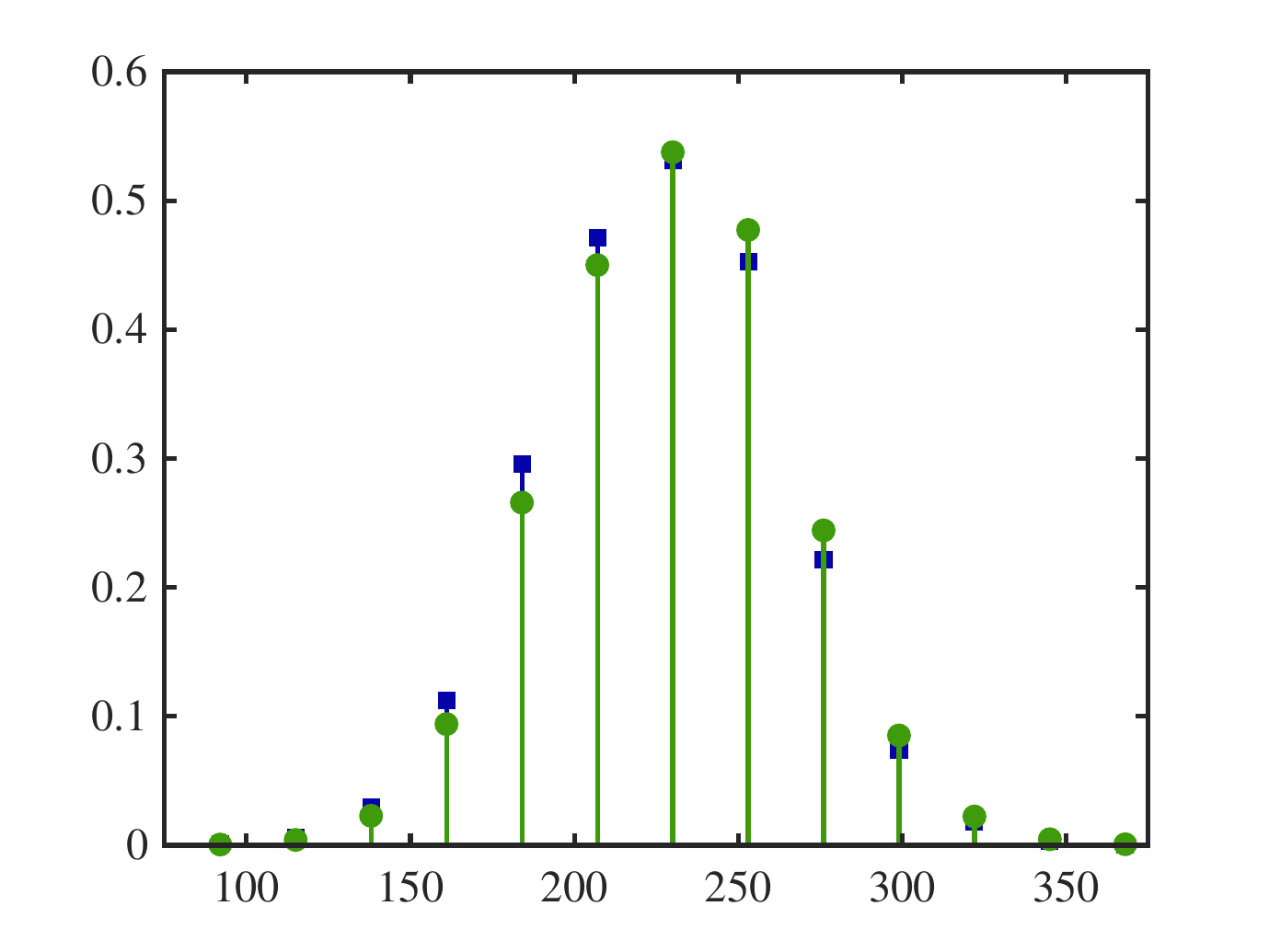}
			\put(52,-1){\htext{\footnotesize $f$ (THz)}}
			\put(3,40){\vtext{\footnotesize $|E_t|$}}
		\end{overpic}
 		\label{Fig:TimeModWeak}
  	}\\
  	\subfloat[$\omega_p=\omega_0/2$, $\Delta_{e}=\Delta_m=0.5$]{%
  		\begin{overpic}[width=0.25\textwidth,grid=false,trim={0cm 0cm 0.5cm 0cm},clip]{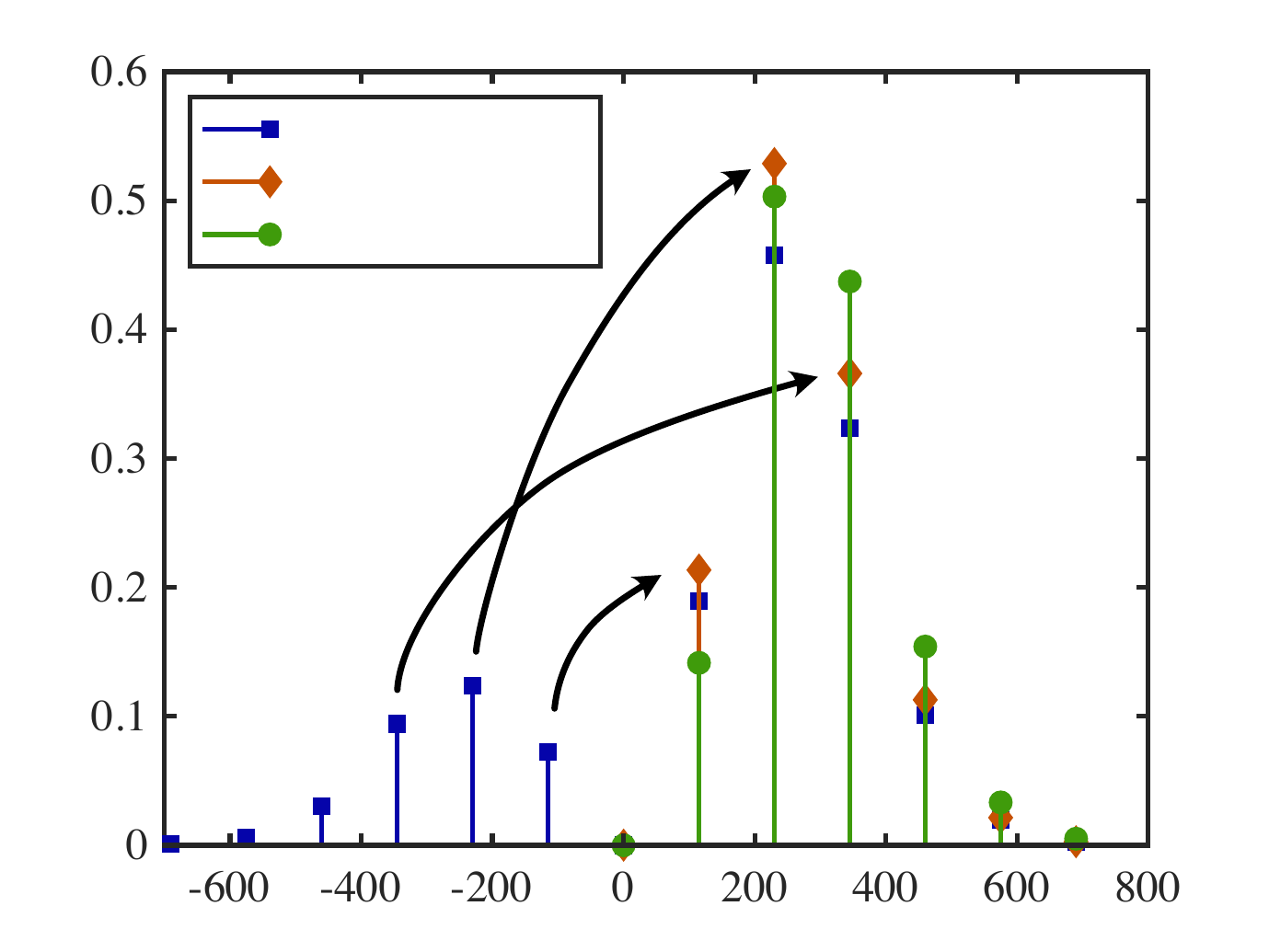}
			\put(52,-1){\htext{\footnotesize $f$ (THz)}}
			\put(3,40){\vtext{\footnotesize $|E_r|$}}
			\put(24,66.4){\tiny Floquet}
			\put(24,61.65){\tiny Floquet (FFT)}
			\put(24,57){\tiny FDTD}
		\end{overpic}
		\begin{overpic}[width=0.25\textwidth,grid=false,trim={0cm 0cm 0.5cm 0cm},clip]{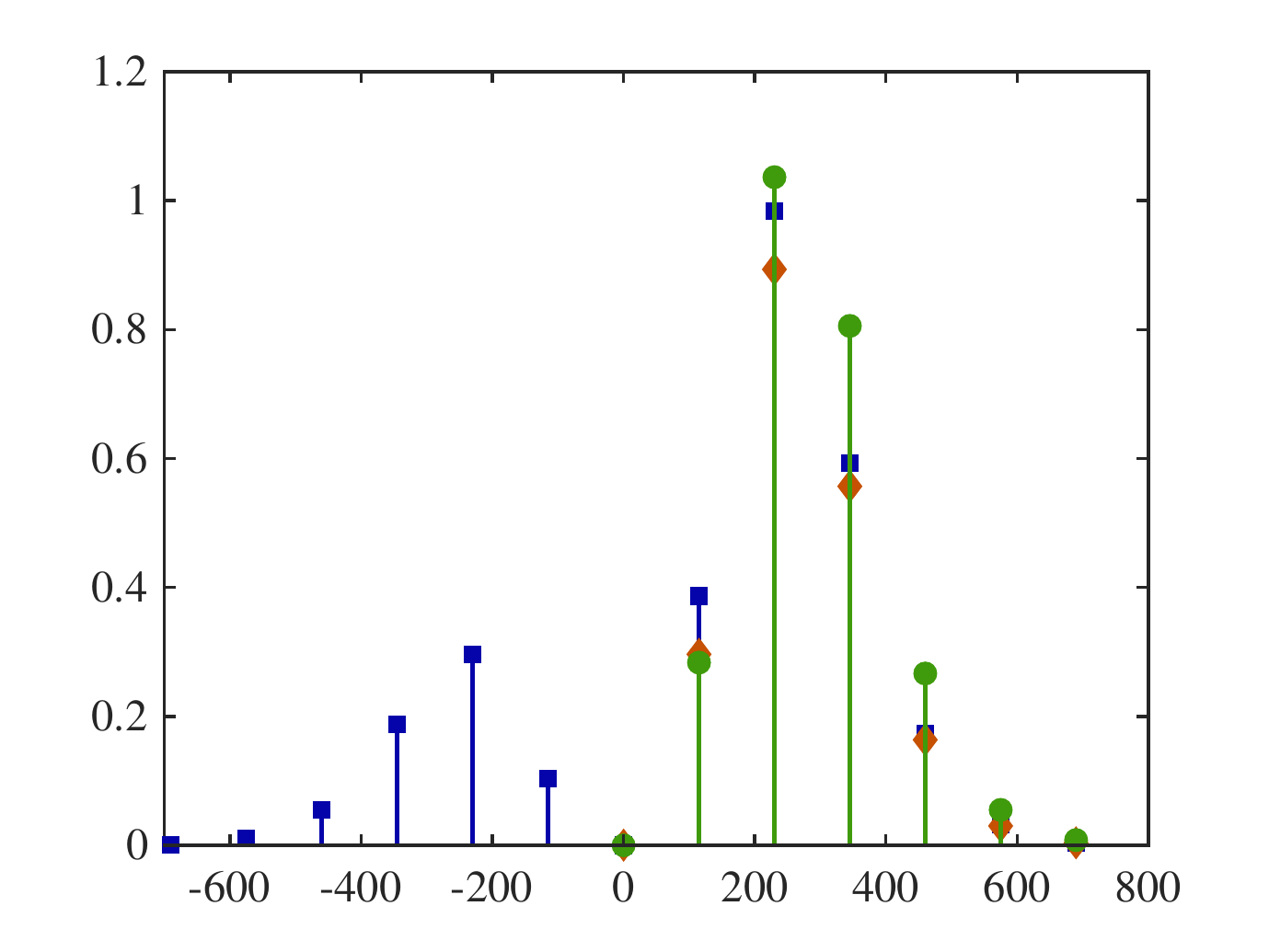}
			\put(52,-1){\htext{\footnotesize $f$ (THz)}}
			\put(3,40){\vtext{\footnotesize $|E_t|$}}
		\end{overpic}
 		\label{Fig:TimeModStrong}
  	}
	\caption{Shows two cases of uniform metasurfaces that are time-modulated with $\omega_{e0,m0}(x)=\omega_{e0,m0}[1+\Delta_{e,m}\cos(\omega_pt){]}$. In (b), the Floquet solution produces negative frequency harmonics, which can be ``folded'' into the positive frequency components, producing the half-sided spectrum in orange diamonds. The form of the modulation is the same as in Fig~\ref{Fig:TimeSteadyState} }
\end{figure}
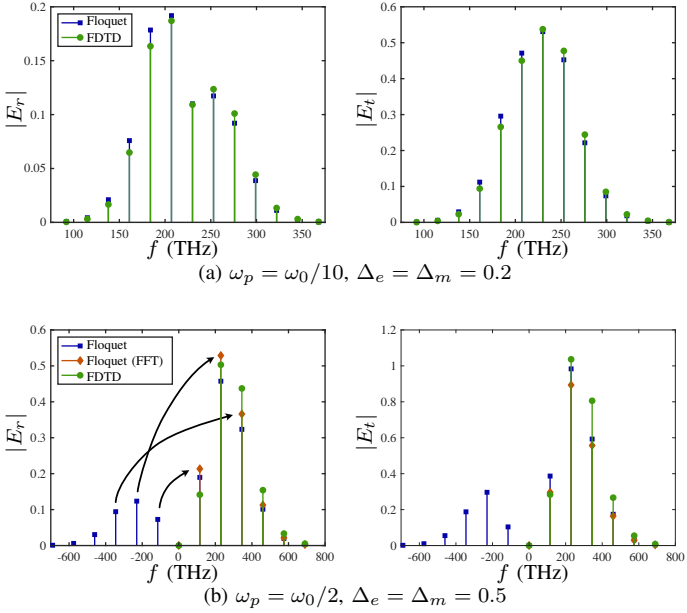

Now, we turn to the temporal modulation of the metasurface. Since the surface is uniform and the incident field is a normal plane wave, this reduces to a 1D problem. To validate the Floquet solution, we use a FDTD technique where the susceptibility is time-variant \cite{Stewart:2018aa}, and run the simulation until a steady-state is achieved. Then a Fourier transform yields the time harmonics that are generated. Fig.~\ref{Fig:TimeSteadyState} shows for instance, the time-domain waveforms obtained using Floquet and FDTD method, for an example of a weak modulation (cosine profile with $\Delta_{e,n}=0.2$), where the waveforms are recorded once the steady-state is reached in FDTD. Fig.~\ref{Fig:TimeModWeak}(a) further shows the corresponding space harmonics showing a good agreement between both Floquet and FDTD solutions.

Next, we consider a stronger modulation ($\Delta_{e,m}=0.5$) while also increasing the pumping frequency to $\omega_p=\omega_0/2$. In this case, harmonics at negative frequencies are excited in the Floquet solution (Fig.~\ref{Fig:TimeModStrong}). By taking a Fourier transform of the time-domain Floquet waveform (orange diamonds), these can be ``flipped'' to positive frequencies; in this case, they combine with positive frequency harmonics because $\omega_p$ is an integer multiple of $\omega_0$. However, even with this taken into account, there is a discrepancy that is observed with the FDTD result.

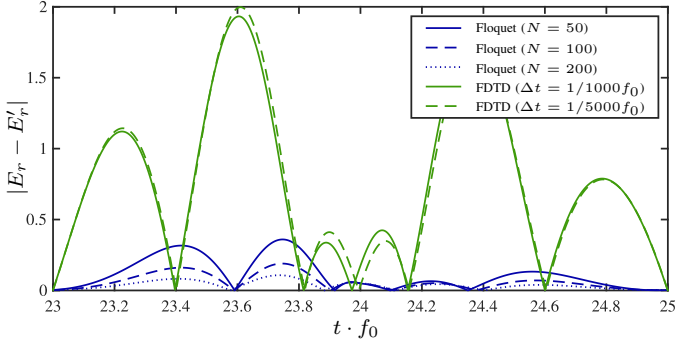
\begin{figure}[h]
	\centering
	\begin{overpic}[width=\columnwidth,grid=false,trim={0cm 0cm 0cm 0cm},clip]{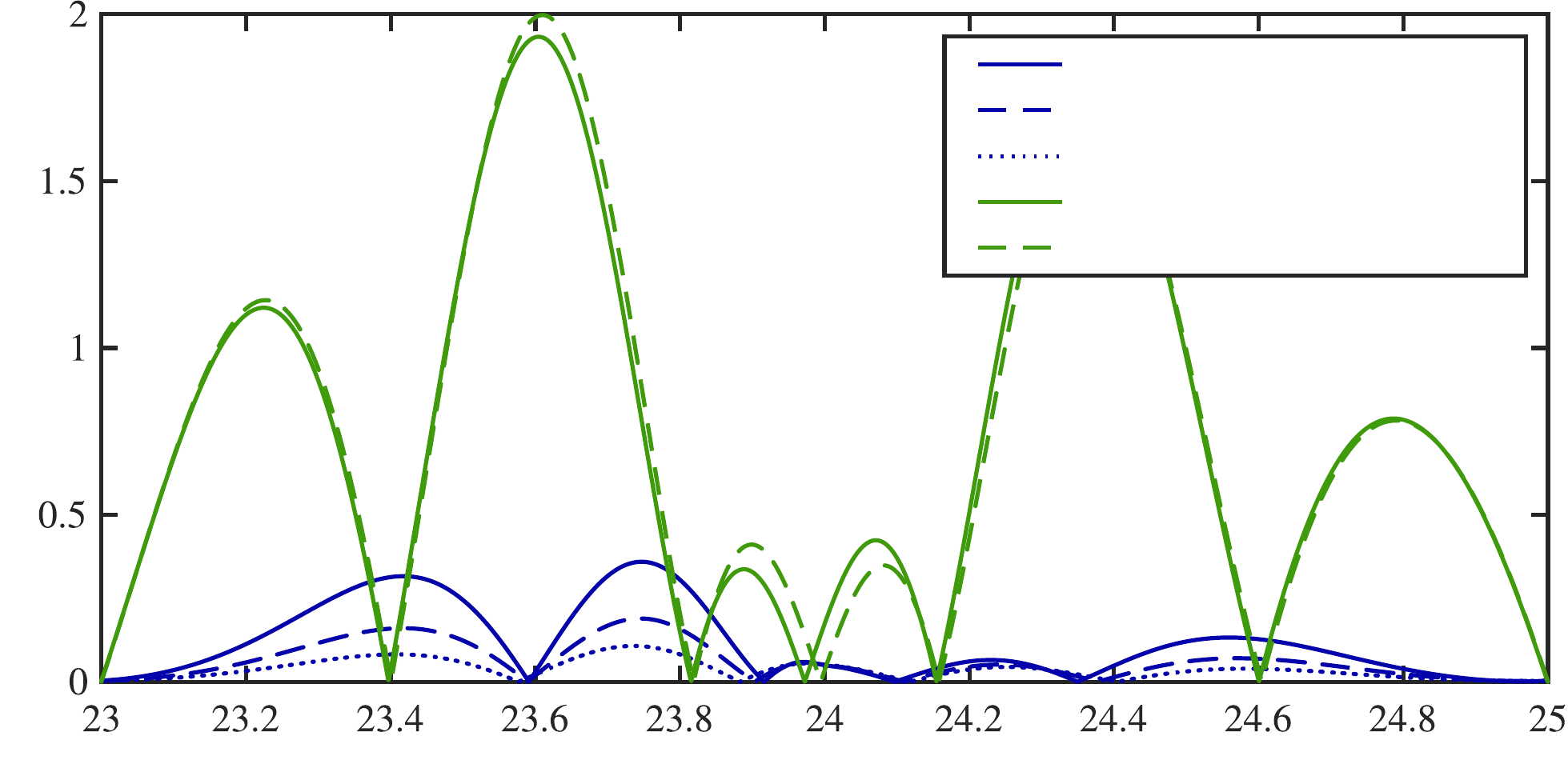}
			\put(1,27){\vtext{\footnotesize $|E_r-E_r'|$}}
			\put(52,0){\htext{\footnotesize $t\cdot f_0$}}
			\put(70,45.3){\tiny Floquet ($N=50$)}
			\put(70,42.2){\tiny Floquet ($N=100$)}
			\put(70,39.1){\tiny Floquet ($N=200$)}
			\put(70,36.3){\tiny FDTD ($\Delta t = 1/1000f_0$)}
			\put(70,33.5){\tiny FDTD ($\Delta t = 1/5000f_0$)}
	\end{overpic}
	\caption{To examine the discrepancy between the FDTD and Floquet results in Fig.~\ref{Fig:TimeModStrong}, the consistency of the solutions was considered. The FDTD discrepancy is larger than the discrepancy of the Floquet result, while the latter improves as the number of harmonics ($2N+1$) is increased.}
	\label{Fig:TimeDiscrepency}
\end{figure}

To determine which result is more accurate, we consider the equations they should satisfy, i.e. \eqref{Eq:GSTCScalar} and \eqref{Eq:Lorentzian}. We can numerically compute the derivatives $dQ_y/dt$ and $d^2Q_y/dt^2$ using the time domain waveforms of FDTD and Floquet methods and substitute into \eqref{Eq:LorentzianQ} to find a new expression $E_{\text{av},y}'$. Similarly, \eqref{Eq:GSTCM} yields a new expression $\Delta E_y'$. Finally, we solve
\begin{subequations}
	\begin{gather}
		E_t' - E_r' = \Delta E_y',\\
		\frac{1}{2}(E_t' + E_r') + E_i = E_{\text{av},y}',
	\end{gather}
\end{subequations}
to find new values $E_r'$ and $E_t'$. If the solution is exact, then we should have $E_r'=E_r$ and $E_t'=E_t$. We carry out this procedure for both the Floquet and FDTD methods, with the disprency $|E_r-E_r'|$ shown in Fig.~\ref{Fig:TimeDiscrepency}. While the FDTD discrepancy changes slowly as time-stepping becomes more fine, the Floquet solution shows convergence as the number of harmonics increases. Furthermore, the Floquet solution has a much smaller discrepancy, indicating that it is the more accurate solution among the two, of the original field equations.

Also, we see that for both methods, the DC electric field harmonics at $\omega_n=0$ are zero (the magnetic field, not shown, is likewise zero). From a physical perspective, a DC $H$ field (or $E$ field) difference can be generated across a boundary due to a static electric current (magnetic current), with the well-known boundary condition derived from Ampere's law (Faraday's law) \cite[p.~76]{Rothwell:2018aa}. In the right hand side of  \eqref{Eq:GSTCScalar}, this corresponds to polarizations that are linearly changing over time, which is in contraction to a periodic solution. Mathematically, this manifests itself in \eqref{Eq:GSTCMSum}, which requires $E_{r0,nm}-E_{r0,nm}=0$ for $\omega_n=0$, while \eqref{Eq:GSTCQSum} requires $E_{r0,nm}+E_{r0,nm}=0$. The solution, of course, is that the DC fields are zero.

\subsection{Space-Time Modulation}\label{Sec:ResultsSpaceTime}


Finally, we consider the general case of space-time modulation. Here, it is convenient to use generalized S-parameters \cite{Caloz:2018aa} to describe the system, where each harmonic in the transmission and reflection regions can be considered a port, for a total of $2(2N+1)(2M+1)$ when the fundamental harmonic $(m,n)=(0,0)$ is normally incident\footnote{This also happens if the fundamental is at $\theta_{00}=\sin^{-1}(m\beta_p)$ with some integer $m$. Otherwise, there will in general be twice as many ports because of a lack of symmetry across the $z$-axis with the incident and reflected field propagation directions.}. We will label the reflection parameter
\begin{subequations}
	\begin{align}
		R_{m_1 n_1}^{m_2n_2} = \left. \frac{E_{r,m_2n_2}}{E_{i,m_in_i}} \right|_{(m_i,n_i)=(-m_1,n_1) }
	\end{align}
which is measured by evaluating \eqref{Eq:EqnExpansions} with the harmonic $(m_in_i)=(-m_1,n_1)$ excited\footnote{We set $m_i=-m_1$ so that the incident ($m_i$) and reflected ($m_2$) wavevectors are parallel, but opposite in direction. Thus, they correspond to the same port in space and frequency.} with a plane wave and the $(m,n)$ port probed. That is, this represents the scattering from port $(m_1,n_1)$ to $(m_2,n_2)$, with $m$ as a spatial index and $n$ as a frequency index that can be used in \eqref{Eq:thetamn} and \eqref{Eq:wn} to find the direction and frequency, respectively. Similarly, the transmission parameter is
	\begin{align}
		T_{m_1n_1}^{m_2n_2} =  \left.\frac{E_{t,m_2,n_2}+E_{i,m_in_i}}{E_{i,m_in_i}} \right|_{(m_i,n_i)=(-m_1,n_1) }
	\end{align}
\end{subequations}

With this convention, Fig.~\ref{Fig:SpaceTimeModeDecoupled} shows a case where the space and time dependencies are decoupled, and the modulation resembles a standing wave. Each pixel represents a scattering parameter with port $(m_1,n_1)=(1,0)$ excited in (a) and $(-1,0)$ excited in (b). One primary interest is whether or not this represents a reciprocal system. One way to approach this is to evaluate if $R_{m_1n_1}^{m_2n_2}=R_{m_2n_2}^{m_1n_1}$ and $T_{m_1n_1}^{m_2n_2}=T_{m_2n_2}^{m_1n_1}$ for all ports combinations \cite{Caloz:2018aa}. For example, we see that $|T_{1,0}^{-1,0}|=|T_{-1,0}^{1,0}|=0.24$ from Fig.~\ref{Fig:SpaceTimeModeDecoupled}, so these ports are reciprocal. 

\newcommand{\modeplot}[3]{%
	\begin{overpic}[#3,grid=false,trim={-0.5cm -0.25cm 0.5cm -0.5cm},clip]{#1}
		\put(49,-1){\htext{\footnotesize $n_2$}}
		\put(3,40){\vtext{\footnotesize $m_2$}}
		\put(49,76){\htext{\footnotesize #2}}
	\end{overpic}
}

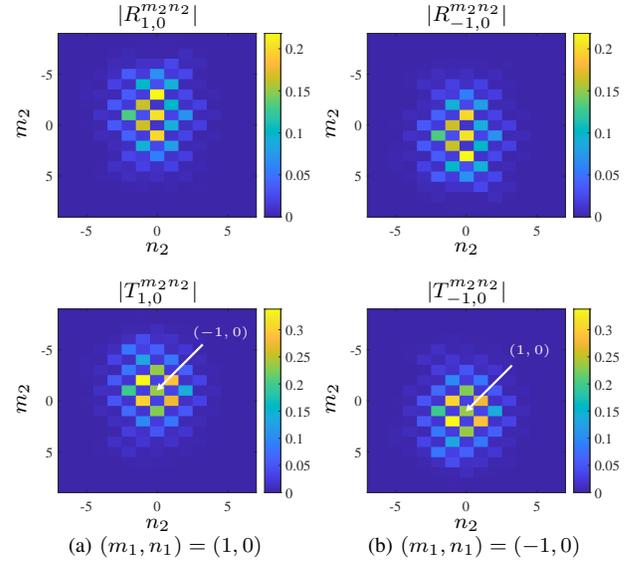
\begin{figure}[h]%
\captionsetup[subfigure]{justification=centering}
 \centering
  	\subfloat[$(m_1,n_1)=(1,0)$]{%
  	\begin{minipage}{0.45\columnwidth}
  		\modeplot{standing_f230_fe230_spacefact5,8_timefact10,0_modl0,20_M12_N12_mi1_ni0_flo_Er0}{$|R_{1,0}^{m_2n_2}|$}{width=\textwidth}\\[0.4cm]
  		\begin{overpic}[width=\textwidth,grid=false,trim={-0.5cm -0.25cm 0.5cm -0.5cm},clip]{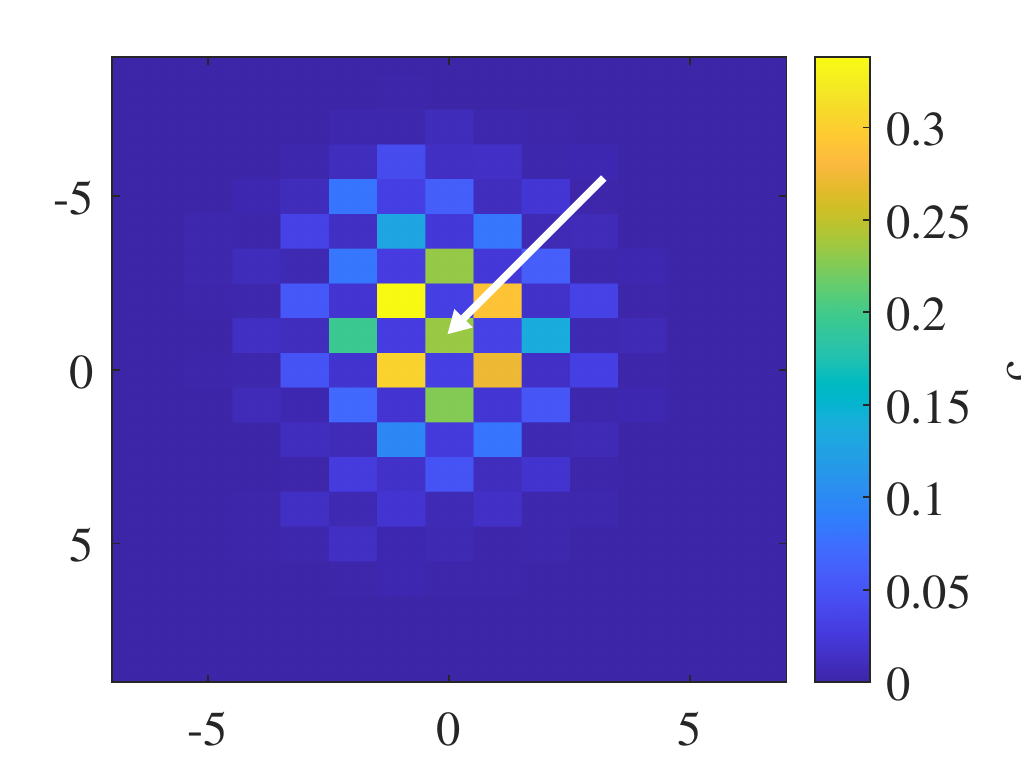}
			\put(49,-1){\htext{\footnotesize $n_2$}}
		    \put(3,40){\vtext{\footnotesize $m_2$}}
			\put(49,76){\htext{\footnotesize $|T_{1,0}^{m_2n_2}|$}}
			\put(70,63){\htext{\color{white} \tiny $(-1,0)$}}
		\end{overpic}
  	\end{minipage}
  	}
  	\subfloat[$(m_1,n_1)=(-1,0)$]{%
  	\begin{minipage}{0.45\columnwidth}
  		\modeplot{standing_f230_fe230_spacefact5,8_timefact10,0_modl0,20_M12_N12_mi-1_ni0_flo_Er0}{$|R_{-1,0}^{m_2n_2}|$}{width=\textwidth}\\[0.4cm]
  		 \begin{overpic}[width=\textwidth,grid=false,trim={-0.5cm -0.25cm 0.5cm -0.5cm},clip]{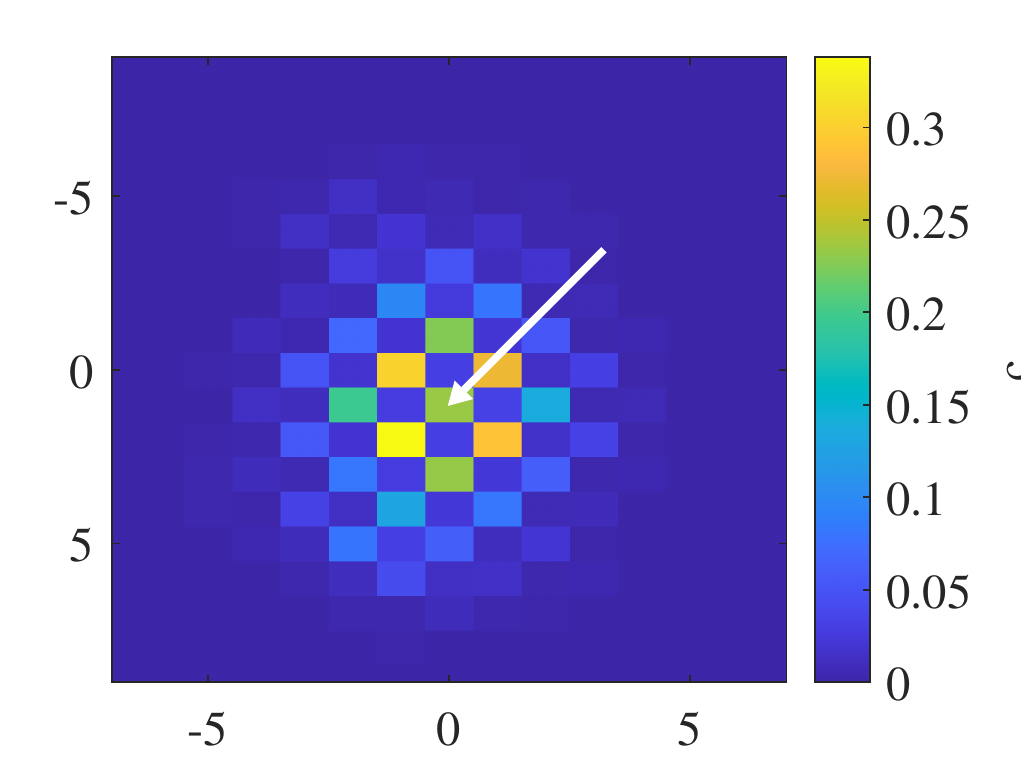}
			\put(49,-1){\htext{\footnotesize $n_2$}}
			\put(3,40){\vtext{\footnotesize $m_2$}}			
			\put(49,76){\htext{\footnotesize $|T_{-1,0}^{m_2n_2}|$}}
			\put(70,57){\htext{\color{white} \tiny $(1,0)$}}
		\end{overpic}
  	\end{minipage}
  	}
 \caption{Reciprocal space-time modulation where $\omega_{e0,m0}(x,t)=\omega_{e0,m0}[1+\Delta_{e,m}\cos(\omega_pt)\cos(\beta_px){]}$ and space and time dependencies are decoupled. Each pixel represents a space-time harmonic, calculated using the Floquet method. For both cases, a plane wave is incident with $f_0=\SI{230}{THz}$, $\Delta_e=\Delta_m=0.2$, $\omega_p=\omega_0/10$, and $\beta_p=k_0/5.76$ so that the first space harmonic is at $\theta=\SI{10}{\degree}$}
 \label{Fig:SpaceTimeModeDecoupled}
\end{figure}

Alternatively, we can consider the Onsager-Casimir relations, which place conditions on the constitutive relations of LTV systems for reciprocity \cite{Caloz:2018aa}\cite{Tretyakov:2002aa}. In the case at hand, these require $\chi_e(v_p)=\chi_e(-v_p)$ and $\chi_m(v_p)=\chi_m(-v_p)$ in order for the system to be reciprocal, where the susceptibilities are a function of $v_p$, which is the velocity of the modulation. Hence, the modulated parameter must also be identical when the direction of modulation is reversed. This is indeed the case for the standing wave modulation, which can be written as the sum of two waves travelling in opposite directions, where switching this sign of velocity is inconsequential ($\omega_{e0,m0}(x,t,v)=\omega_{e0,m0}(x,t,-v)$). This can be intuitively understood as follows: the surface ``appears'' the same to an incident wave regardless of which side of the surface it approaches from.

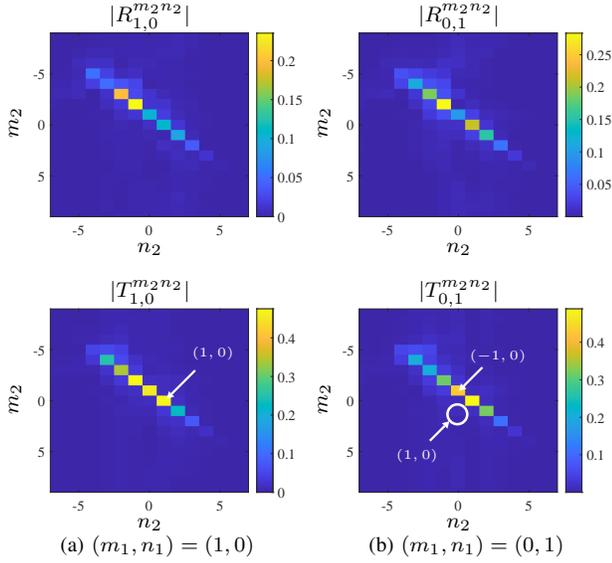
\begin{figure}[h]%
 \centering
  	\subfloat[$(m_1,n_1)=(1,0)$]{%
  	\begin{minipage}{0.45\columnwidth}
  		\modeplot{travelling_f230_fe230_spacefact5,8_timefact10,0_modl0,20_M12_N12_mi1_ni0_flo_Er0}{$|R_{1,0}^{m_2n_2}|$}{width=\textwidth}\\[0.4cm]
  		\begin{overpic}[width=\textwidth,grid=false,trim={-0.5cm -0.25cm 0.5cm -0.5cm},clip]{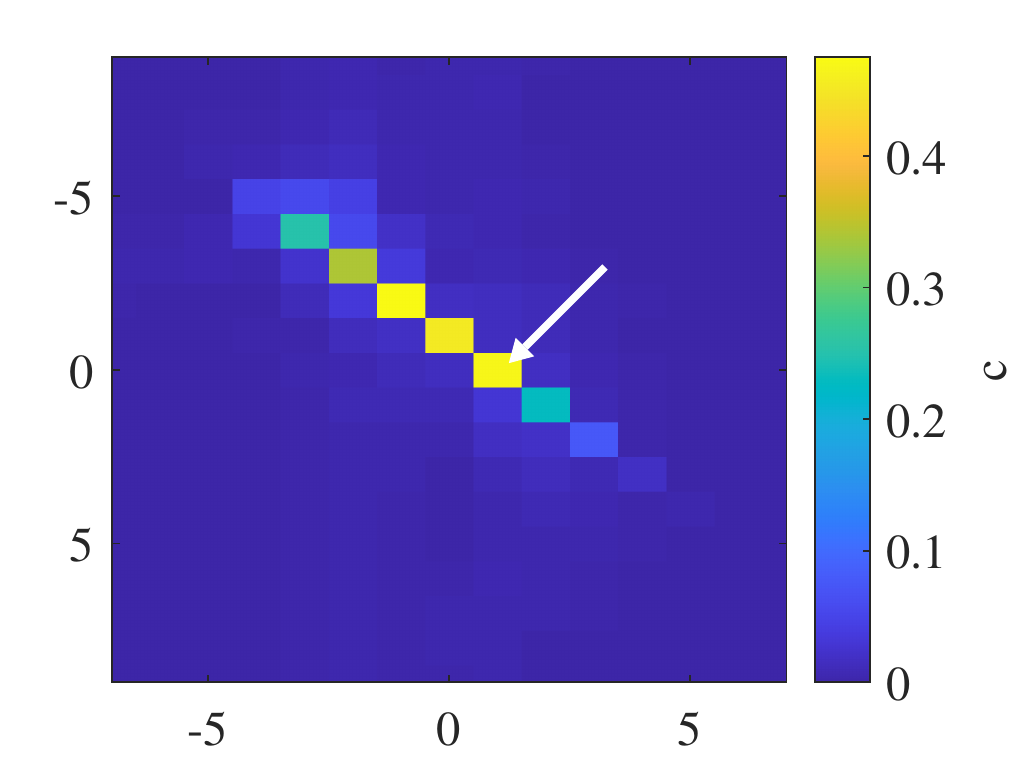}
			\put(49,-1){\htext{\footnotesize $n_2$}}
			\put(3,40){\vtext{\footnotesize $m_2$}}
			\put(49,76){\htext{\footnotesize $|T_{1,0}^{m_2n_2}|$}}
			\put(70,56){\htext{\color{white} \tiny $(1,0)$}}
		\end{overpic}
  	\end{minipage}
  	}
  	\subfloat[$(m_1,n_1)=(0,1)$]{%
  	\begin{minipage}{0.45\columnwidth}
  		\modeplot{travelling_f253_fe230_spacefact5,8_timefact10,0_modl0,20_M12_N12_mi0_ni1_flo_Er0}{$|R_{0,1}^{m_2n_2}|$}{width=\textwidth}\\[0.4cm]
  		\begin{overpic}[width=\textwidth,grid=false,trim={-0.5cm -0.25cm 0.5cm -0.5cm},clip]{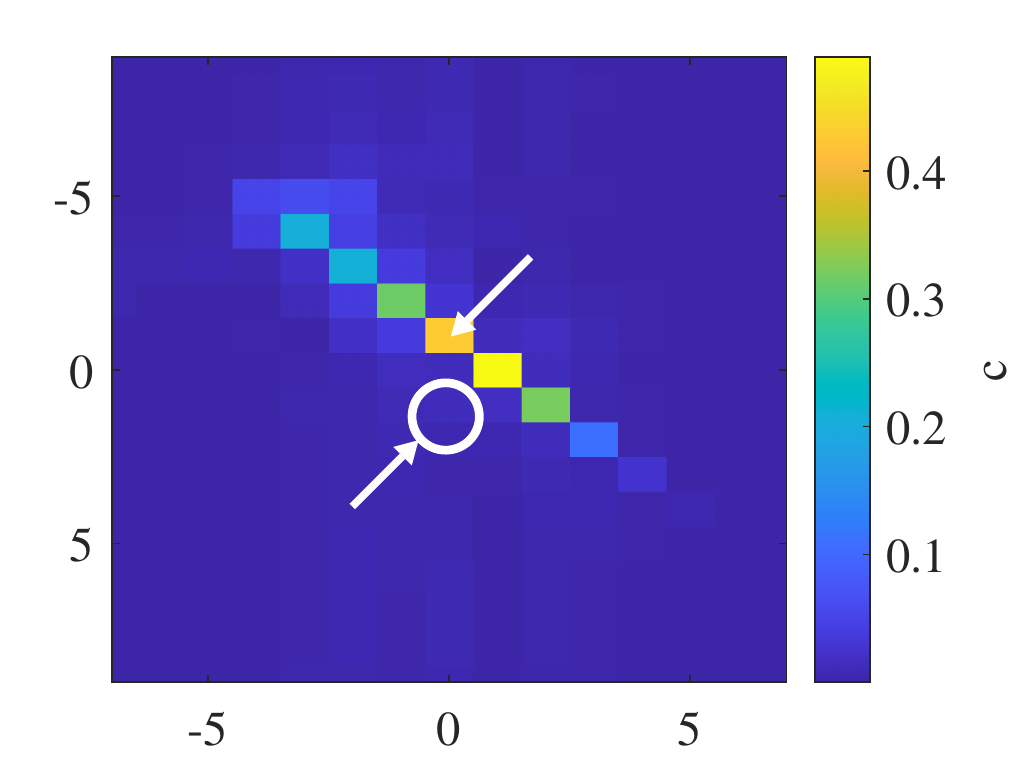}
			\put(49,-1){\htext{\footnotesize $n_2$}}
			\put(3,40){\vtext{\footnotesize $m_2$}}
			\put(49,76){\htext{\footnotesize $|T_{0,1}^{m_2n_2}|$}}
			\put(62,55){\htext{\color{white} \tiny $(-1,0)$}}
			\put(35,22){\htext{\color{white} \tiny $(1,0)$}}
		\end{overpic}
  	\end{minipage}
  	}
 \caption{Non-reciprocal space-time modulated metasurface where $\omega_{e0,m0}(x,t)=\omega_{e0,m0}[1+\Delta_{e,m}\cos(\omega_pt-\beta_px){]}$ and space and time dependencies are coupled. Aside from the form of the modulation, the parameters are the same as in Fig.~\ref{Fig:SpaceTimeModeDecoupled}.}
 \label{Fig:SpaceTimeModCoupled}
\end{figure}

To break reciprocity, we consider a case where the space and time modulations are coupled in the form of a wave travelling along the surface in the $+x$ direction. The scattering parameters are in Fig~\ref{Fig:SpaceTimeModCoupled}, where in (a), we see that exciting the $(1,0)$ port we observe an up-converted transmitted harmonic at $(0,1)$ with $|T_{1,0}^{0,1}|=0.47$. Exciting this port in hand, we find $|T_{0,1}^{1,0}|\approx 0$! (Instead port $(-1,0)$ is excited.) Thus the system is non-reciprocal. Of course, the Onsager-Casimir relations are not satisfied in this case, since the direction of modulation is critical.

Extending the analysis from a plane wave to a more general incident field, and at the same time visually demonstrating the non-reciprocity, Fig.~\ref{Fig:SpaceTimeGaussian} shows the fields of several frequency harmonics when a Gaussian beam is launched at the metasurface in the same two experiments. In the first case (a), a beam incident on the $(1,0)$ port (\SI{10}{\degree} at $1.0\omega_0$) has a normally transmitted harmonic that is up-converted ($1.1\omega_0$), corresponding to port $(0,1)$. If we in turn excite this port in (b), we do not find find a transmitted harmonic in the direction of the first incident beam at $\omega_0$. This harmonic at $1.0\omega_0$ is instead directed at an angle \SI{-10}{\degree}, clearly demonstrating the nonreciprocal nature of the surface.

\newcommand{\fieldplot}[3]{%
	\begin{overpic}[#3,grid=false,trim={1.2cm -0.5cm 1.5cm -0.5cm},clip]{#1}
		\put(46,-1){\htext{\footnotesize $n'$}}
		\put(3,49){\vtext{\footnotesize $m'$}}
		\put(46,91){\htext{\footnotesize #2}}
	\end{overpic}
}

\newcommand{\fieldplotcb}[3]{%
	\begin{overpic}[#3,grid=false,trim={0.5cm -0.5cm 1.5cm -0.5cm},clip]{#1}
		\put(45.5,-1){\htext{\footnotesize $n'$}}
		\put(3,49){\vtext{\footnotesize $m'$}}
		\put(45.5,91){\htext{\footnotesize #2}}
	\end{overpic}
}

\begin{figure}[h]
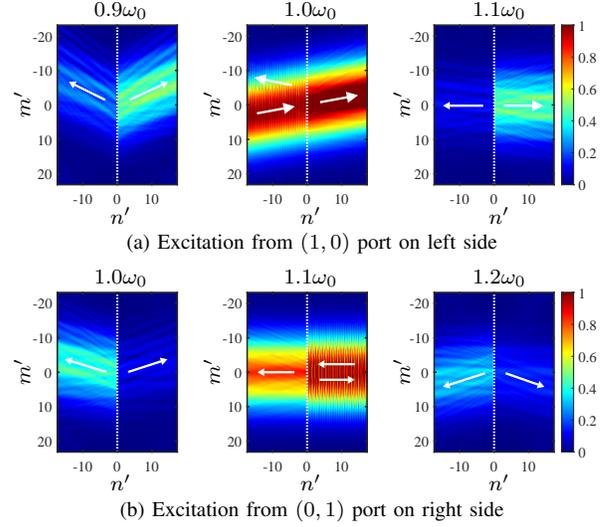
%
 \centering
  	\subfloat[Excitation from $(1,0)$ port on left side]{%
  	\fieldplot{gaussian_-10deg_f230_fe230_spacefact5,8_timefact10,0_modl0,20_M8_N8_flo_Etot_n-1_2}{$0.9\omega_0$}{height=3cm}%
  	\fieldplot{gaussian_-10deg_f230_fe230_spacefact5,8_timefact10,0_modl0,20_M8_N8_flo_Etot_n0_2}{$1.0\omega_0$}{height=3cm}%
  	\fieldplotcb{gaussian_cb_-10deg_f230_fe230_spacefact5,8_timefact10,0_modl0,20_M8_N8_flo_Etot_n1_2}{$1.1\omega_0$}{height=3cm}}\\[0.1cm]
  	\subfloat[Excitation from $(0,1)$ port on right side]{%
  	\fieldplot{gaussian_0deg_f2,530000e+02_fe230_spacefact5,8_timefact10,0_modl0,20_M8_N8_flo_Etot_n-1_2}{$1.0\omega_0$}{height=3cm}%
  	\fieldplot{gaussian_0deg_f2,530000e+02_fe230_spacefact5,8_timefact10,0_modl0,20_M8_N8_flo_Etot_n0_2}{$1.1\omega_0$}{height=3cm}%
  	\fieldplotcb{gaussian_cb_0deg_f2,530000e+02_fe230_spacefact5,8_timefact10,0_modl0,20_M8_N8_flo_Etot_n1_2}{$1.2\omega_0$}{height=3cm}}
 \caption{A \SI{230}{THz} gaussian beam with a waist $10\lambda_0$ is incident on a surface with a modulation $\omega_{e0,m0}(x,t)=\omega_{e0,m0}[1+\Delta_{e,m}\cos(\omega_pt-\beta_px){]}$ and parameters identical to Fig.~\ref{Fig:SpaceTimeModCoupled}. }
 \label{Fig:SpaceTimeGaussian}
\end{figure}
\section{Conclusions}\label{Sec:Conclusion}

A rigorous semi-analytical Floquet analysis has been presented for a zero-thickness space-time modulated Huygens' metasurface using GSTCs to model and determine the strengths of the new harmonic components of the scattered fields. We have accounted for the dispersion inherent to the static metasurface using physically-motived Lorentzian susceptibilities, with parameters that are modulated in space and time. These parameters ($\omega_0$, $\omega_p$, and $\alpha$) can take on arbitrary periodic profiles in space and time, for both the electric and magnetic susceptibilities. The validity of the method has been established with comparison to FDFD simulations for pure-space modulation and FDTD simulations for pure-time modulation. Finally, two cases of space time modulation were presented: a standing wave perturbation which was found to be reciprocal and a traveling wave perturbation that breaks Lorentz reciprocity. The proposed method is fast, simple, and versatile, and is expected to be a useful tool for designing general periodic and non-reciprocal metasurfaces.

%

\appendices

\section{Matrix Formulation}\label{Sec:AppendixFormulation}
We can arrange \eqref{Eq:EqnExpansions} into a finite matrix to create a tractable computational problem. Since there are two indexes, $m$ and $n$, it is convenient map the permutations to an index $p$, which we can denote the space/time indexes as $m(p)$ and $n(p)$. Having done this, the unknown harmonic amplitudes can be written as column vectors, i.e.
\begin{align}
	\textbf{E}_t = \begin{bmatrix}
		E_{t,m(0)n(0)}, E_{t,m(1)n(1)},  \cdots, E_{t,m(p_\text{max})n(p_\text{max})}
	\end{bmatrix}_,^T
\end{align}
and likewise for $\textbf{E}_r$, $\textbf{Q}$, and $\textbf{M}$. Now, \eqref{Eq:EqnExpansions} becomes

\begin{align}
    \begin{bmatrix}
        \mathbf{X}_{e1} & \nullv & -\mathbf{X}_{e2} & -\mathbf{X}_{e2}\\
        \nullv & \mathbf{X}_{m1} & \mathbf{X}_{m2} & -\mathbf{X}_{m2}\\
        \nullv & \mathbf{W}_1 & -\mathbf{I} & \mathbf{I}\\
        \mathbf{W}_2 & \nullv & \mathbf{\Theta} & \mathbf{\Theta}
    \end{bmatrix}%
    \begin{bmatrix}
        \mathbf{Q}\\
        \mathbf{M}\\
        \mathbf{E}_t\\
        \mathbf{E}_r
    \end{bmatrix}=%
    \begin{bmatrix}
        \mathbf{X}_{e3}\\
        -\mathbf{X}_{m3}\\
        \nullv \\
        \nullv \\
    \end{bmatrix}
\end{align}
where 
\begin{subequations}
	\begin{gather*}
		\mathbf{X}_{e3}(p) = \omega_{ep,\Delta (p_i,p)},\\
	    \mathbf{X}_{m3}(p) = \frac{\omega_{mp,\Delta (p_i,p)}}{\eta_0},
	\end{gather*}\\[-1cm]
	\begin{multline*}
		\mathbf{X}_{e1}(p_1,p_2) = \omega_{e0,\Delta (p_1,p_2)} - \delta(p_1-p_2)\omega_{n(p_1)}^2 \\+ j\omega_{n(p_2)} \alpha_{e,\Delta (p_1,p_2)},
	\end{multline*}\\[-1cm]
	\begin{gather*}
	    \mathbf{X}_{e2}(p_1,p_2) = \frac{\omega_{ep,\Delta (p_1,p_2)}}{2},
	\end{gather*}\\[-1cm]
	\begin{multline*}
	    \mathbf{X}_{m1}(p_1,p_2) = \omega_{m0,\Delta (p_1,p_2)} - \delta(p_2-p_1)\omega_{n(p_1)}^2 \\
	    + j\omega_{n(p_2)} \alpha_{m,\Delta (p_1,p_2)},
	\end{multline*}\\[-1cm]
	\begin{gather*}
	    \mathbf{X}_{m2}(p_1,p_2) = \frac{\omega_{mp,\Delta (p_1,p_2)}\cos\theta_{m(p_2),n(p_2)}}{2\eta_0},
	\end{gather*}	
\end{subequations}
where, for example, $\mathbf{X}_{e1}(p_1,p_2)$ is the element at row $p_1$ and column $p_2$, $\Delta (p_1,p_2)=(m(p_2)-m(p_1),n(p_2)-n(p_1))$, and $p_i$ is the index of the incident plane wave harmonic. The remaining matrices are zero except when $p_1=p_2$:
\begin{subequations}
	\begin{gather*}
	    \mathbf{W}_{1}(p,p) = j\mu_0\omega_{n(p)}\\
	    \mathbf{W}_{2}(p,p) = \frac{j\omega_{n(p)}}{c_0\cos\theta_{i}}\\
	    \mathbf{I}(p,p) = 1\\
	    \mathbf{\Theta}(p,p) = \frac{\cos\theta_{m(p),n(p)}}{\cos\theta_{i}}
	\end{gather*}	
\end{subequations}

\bibliographystyle{IEEEtran}
\bibliography{stp,stp2}

\begin{thebibliography}{10}
\providecommand{\url}[1]{#1}
\csname url@samestyle\endcsname
\providecommand{\newblock}{\relax}
\providecommand{\bibinfo}[2]{#2}
\providecommand{\BIBentrySTDinterwordspacing}{\spaceskip=0pt\relax}
\providecommand{\BIBentryALTinterwordstretchfactor}{4}
\providecommand{\BIBentryALTinterwordspacing}{\spaceskip=\fontdimen2\font plus
\BIBentryALTinterwordstretchfactor\fontdimen3\font minus
  \fontdimen4\font\relax}
\providecommand{\BIBforeignlanguage}[2]{{%
\expandafter\ifx\csname l@#1\endcsname\relax
\typeout{** WARNING: IEEEtran.bst: No hyphenation pattern has been}%
\typeout{** loaded for the language `#1'. Using the pattern for}%
\typeout{** the default language instead.}%
\else
\language=\csname l@#1\endcsname
\fi
#2}}
\providecommand{\BIBdecl}{\relax}
\BIBdecl

\bibitem{Cullen:1958aa}
A.~L. Cullen, ``A travelling-wave parametric amplifier,'' \emph{Nature}, vol.
  181, no. 4605, pp. 332--332, 1958.

\bibitem{Cassedy:1963aa}
E.~S. Cassedy and A.~A. Oliner, ``Dispersion relations in time-space periodic
  media: Part {I}---stable interactions,'' \emph{Proceedings of the IEEE},
  vol.~51, no.~10, pp. 1342--1359, 1963.

\bibitem{Chen:2016aa}
H.-T. Chen, A.~J. Taylor, and N.~Yu, ``A review of metasurfaces: physics and
  applications,'' \emph{Rep. Prog. Phys.}, vol.~79, no.~7, p. 076401, 2016.

\bibitem{Genevet:2015aa}
P.~Genevet and F.~Capasso, ``Holographic optical metasurfaces: a review of
  current progress,'' \emph{Rep. Prog. Phys.}, vol.~78, no.~2, p. 024401, 2015.

\bibitem{Adam:2002aa}
J.~D. Adam, L.~E. Davis, G.~F. Dionne, E.~F. Schloemann, and S.~N. Stitzer,
  ``Ferrite devices and materials,'' \emph{IEEE Trans. Microw. Theory Techn.},
  vol.~50, no.~3, pp. 721--737, 2002.

\bibitem{Shi:2015aa}
Y.~Shi, Z.~Yu, and S.~Fan, ``Limitations of nonlinear optical isolators due to
  dynamic reciprocity,'' \emph{Nature Photon.}, vol.~9, no.~6, pp. 388--392,
  2015.

\bibitem{Caloz:2020aa}
C.~Caloz and Z.~Deck-L{\'e}ger, ``Spacetime metamaterials---part {I}: General
  concepts,'' \emph{{IEEE} Trans. Antennas Propag.}, vol.~68, no.~3, pp.
  1569--1582, 2020.

\bibitem{Caloz:2020ab}
------, ``Spacetime metamaterials---part {II}: Theory and applications,''
  \emph{{IEEE} Trans. Antennas Propag.}, vol.~68, no.~3, pp. 1583--1598, 2020.

\bibitem{Wang:2020aa}
X.~Wang, A.~D{\'\i}az-Rubio, H.~Li, S.~A. Tretyakov, and A.~Al{\`u}, ``Theory
  and design of multifunctional space-time metasurfaces,'' \emph{Phys. Rev.
  Appl.}, vol.~13, no.~4, p. 044040, 04 2020.

\bibitem{Hadad:2015aa}
Y.~Hadad, D.~L. Sounas, and A.~Alu, ``Space-time gradient metasurfaces,''
  \emph{Phys. Rev. B}, vol.~92, no.~10, p. 100304, 09 2015.

\bibitem{Taravati:2017aa}
S.~Taravati, N.~Chamanara, and C.~Caloz, ``Nonreciprocal electromagnetic
  scattering from a periodically space-time modulated slab and application to a
  quasisonic isolator,'' \emph{Phys. Rev. B}, vol.~96, no.~16, p. 165144, 10
  2017.

\bibitem{Taravati:2020aa}
S.~Taravati and A.~A. Kishk, ``Space-time modulation: Principles and
  applications,'' \emph{{IEEE} Microw. Mag.}, vol.~21, no.~4, pp. 30--56, 2020.

\bibitem{Ramaccia:2020aa}
D.~Ramaccia, D.~L. Sounas, A.~Al{\`u}, A.~Toscano, and F.~Bilotti,
  ``Phase-induced frequency conversion and {D}oppler effect with time-modulated
  metasurfaces,'' \emph{{IEEE} Trans. Antennas Propag.}, vol.~68, no.~3, pp.
  1607--1617, 2020.

\bibitem{Taravati:2019aa}
S.~Taravati and G.~V. Eleftheriades, ``Generalized space-time-periodic
  diffraction gratings: Theory and applications,'' \emph{Phys. Rev. Appl.},
  vol.~12, no.~2, p. 024026, 08 2019.

\bibitem{Elliptical_DMS}
A.~Arbabi, Y.~Horie, M.~Bagheri, and A.~Faraon, ``Complete control of
  polarization and phase of light with high efficiency and sub-wavelength
  spatial resolution,'' \emph{arXiv:1411.1494}, pp. 4308--4315, Nov 2014.

\bibitem{GeneralizedRefraction}
N.~Yu, P.~Genevet, M.~A. Kats, F.~Aieta, J.-P. Tetienne, F.~Capasso, and
  Z.~Gaburro, ``Light propagation with phase discontinuities: Generalized laws
  of reflection and refraction,'' \emph{Science}, vol. 334, no. 6054, pp.
  333--337, 2011.

\bibitem{meta3}
N.~Yu and F.~Capasso, ``Flat optics with designer metasurfaces,'' \emph{Nat.
  Materials}, vol.~13, April 2014.

\bibitem{Kerker_Scattering}
M.~Kerker, \emph{The Scattering of Light and Other Electromagnetic
  Radiation}.\hskip 1em plus 0.5em minus 0.4em\relax Academic Press, New York,
  1969.

\bibitem{Kivshar_Alldielectric}
M.~Decker, I.~Staude, M.~Falkner, J.~Dominguez, D.~N. Neshev, I.~Brener,
  T.~Pertsch, and Y.~S. Kivshar, ``High-efficiency dielectric {H}uygens'
  surfaces,'' \emph{Adv. Opt. Mater.}, vol.~3, no.~6, pp. 813--820, 2015.

\bibitem{AllDieelctricMTMS}
S.~Jahani and Z.~Jacob, ``All-dielectric metamaterials,'' \emph{Nat.
  Nanotech.}, vol.~2, no.~11, pp. 23--36, Jan 2016.

\bibitem{Grbic_Metasurfaces}
C.~Pfeiffer and A.~Grbic, ``Metamaterial {H}uygens' surfaces: Tailoring wave
  fronts with reflectionless sheets,'' \emph{Phys. Rev. Lett.}, vol. 110, p.
  197401, May 2013.

\bibitem{HuygenBook_Eleftheriades}
J.~G. Webster, \emph{Controlling Electromagnetic Wavefronts Using {H}uygens'
  Metasurfaces}.\hskip 1em plus 0.5em minus 0.4em\relax John Wiley \& Sons,
  Inc., 1999.

\bibitem{Smy:2017aa}
T.~J. Smy and S.~Gupta, ``Finite-difference modeling of broadband {H}uygens'
  metasurfaces based on generalized sheet transition conditions,'' \emph{{IEEE}
  Trans. Antennas Propag.}, vol.~65, no.~5, pp. 2566--2577, 2017.

\bibitem{Stewart:2018aa}
S.~A. Stewart, T.~J. Smy, and S.~Gupta, ``Finite-difference time-domain
  modeling of space--time-modulated metasurfaces,'' \emph{{IEEE} Trans.
  Antennas Propag.}, vol.~66, no.~1, pp. 281--292, 2018.

\bibitem{Smy:2020aa}
T.~J. Smy, S.~A. Stewart, J.~G.~N. Rahmeier, and S.~Gupta, ``{FDTD} simulation
  of dispersive metasurfaces with {L}orentzian surface susceptibilities,''
  \emph{{IEEE} Access}, vol.~8, pp. 83\,027--83\,040, 2020.

\bibitem{Kuester:2003aa}
E.~F. Kuester, M.~A. Mohamed, M.~Piket-May, and C.~L. Holloway, ``Averaged
  transition conditions for electromagnetic fields at a metafilm,''
  \emph{{IEEE} Trans. Antennas Propag.}, vol.~51, no.~10, pp. 2641--2651, 2003.

\bibitem{Vahabzadeh:2018ab}
Y.~Vahabzadeh, N.~Chamanara, K.~Achouri, and C.~Caloz, ``Computational analysis
  of metasurfaces,'' \emph{{IEEE} J. Multiscale Multiphys. Comput. Techn.},
  vol.~3, pp. 37--49, 2018.

\bibitem{Stewart:2019aa}
S.~A. Stewart, S.~Moslemi-Tabrizi, T.~J. Smy, and S.~Gupta, ``Scattering field
  solutions of metasurfaces based on the boundary element method for
  interconnected regions in 2-d,'' \emph{{IEEE} Trans. Antennas Propag.},
  vol.~67, no.~12, pp. 7487--7495, 2019.

\bibitem{Inampudi:2019aa}
S.~Inampudi, M.~M. Salary, S.~Jafar-Zanjani, and H.~Mosallaei, ``Rigorous
  space-time coupled-wave analysis for patterned surfaces with temporal
  permittivity modulation,'' \emph{Opt. Mater. Express}, vol.~9, no.~1, pp.
  162--182, 2019.

\bibitem{Lindell:1994aa}
I.~V. Lindell, A.~H. Sihvola, S.~Tretyakov, and A.~Viitar,
  \emph{Electromagnetic Waves in Chiral and Bi-isotropic Media}.\hskip 1em plus
  0.5em minus 0.4em\relax Norwood, MA, USA: Artech House, 1994.

\bibitem{Achouri:2015aa}
K.~Achouri, M.~A. Salem, and C.~Caloz, ``General metasurface synthesis based on
  susceptibility tensors,'' \emph{{IEEE} Trans. Antennas Propag.}, vol.~63,
  no.~7, pp. 2977--2991, 2015.

\bibitem{Albooyeh:2017aa}
M.~Albooyeh, H.~Kazemi, F.~Capolino, D.~. Kwon, and S.~A. Tretyakov, ``Normal
  vs tangential polarizations in metasurfaces,'' in \emph{{IEEE} Int. Symp.
  Antennas Propag. \& {USNC/URSI} Nat. Radio Sci. Meeting}, 2017, pp.
  1707--1708.

\bibitem{Selvanayagam:2013aa}
M.~Selvanayagam and G.~V. Eleftheriades, ``Discontinuous electromagnetic fields
  using orthogonal electric and magnetic currents for wavefront manipulation,''
  \emph{Opt. Express}, vol.~21, no.~12, pp. 14\,409--14\,429, 2013.

\bibitem{Lathi:2018aa}
B.~P. Lathi and R.~Green, \emph{Linear Systems and Signals}, 3rd~ed.\hskip 1em
  plus 0.5em minus 0.4em\relax New York, NY, USA: Oxford Univ. Press,, 2018, p.
  170.

\bibitem{Idemen:1987aa}
M.~Idemen and A.~H. Serbest, ``Boundary conditions of the electromagnetic
  field,'' \emph{Electronics Lett.}, vol.~23, no.~13, pp. 704--705, 1987.

\bibitem{Claasen:1982aa}
T.~Claasen and W.~Mecklenbrauker, ``On stationary linear time-varying
  systems,'' \emph{{IEEE} Trans. Circuits Syst.}, vol.~29, no.~3, pp. 169--184,
  1982.

\bibitem{Franks:1969aa}
L.~E. Franks, \emph{Signal Theory}.\hskip 1em plus 0.5em minus 0.4em\relax
  Englewood Cliffs, NJ, USA: Prentice-Hall, 1969.

\bibitem{Zadeh:1950aa}
L.~A. Zadeh, ``Frequency analysis of variable networks,'' \emph{Proc. {IRE}},
  vol.~38, no.~3, pp. 291--299, 1950.

\bibitem{Rothwell:2018aa}
E.~Rothwell and M.~Cloud, \emph{Electromagnetics}, 3rd~ed.\hskip 1em plus 0.5em
  minus 0.4em\relax Boca Raton, FL, USA: {CRC} Press, 2018.

\bibitem{Ding:2017aa}
F.~Ding, A.~Pors, and S.~I. Bozhevolnyi, ``Gradient metasurfaces: a review of
  fundamentals and applications,'' vol.~81, no.~2, p. 026401, 2017.

\bibitem{Caloz:2018aa}
C.~Caloz, A.~Al{\`u}, S.~Tretyakov, D.~Sounas, K.~Achouri, and Z.-L.
  Deck-L{\'e}ger, ``Electromagnetic nonreciprocity,'' \emph{Physical Rev.
  Appl.}, vol.~10, no.~4, p. 047001, 10 2018.

\bibitem{Tretyakov:2002aa}
S.~Tretyakov, A.~Sihvola, and B.~Jancewicz, ``{Onsager-Casimir} principle and
  the constitutive relations of bi-anisotropic media,'' \emph{J. Electromagn.
  Waves Appl.}, vol.~16, no.~4, pp. 573--587, 2002.

\end{thebibliography}

\end{document}